\documentclass[showpacs,amssymb,twocolumn,floatfix,pre,aps,notitlepage]{revtex4-1}

\usepackage{amssymb,amsfonts,amsmath,bm}
\usepackage{graphics,graphicx}
\usepackage{color}

\newcommand{\emma}[1]{ #1}

\begin{document}

\title{Electric double layers with surface charge modulations: Novel exact
Poisson-Boltzmann solutions}

\author{Ladislav \v{S}amaj$^{1}$} 
\author{Emmanuel Trizac$^2$} 

\affiliation{
$^1$Institute of Physics, Slovak Academy of Sciences, Bratislava, Slovakia\\
$^2$LPTMS, CNRS, Univ. Paris-Sud, Universit\'e Paris-Saclay, 91405 Orsay,
France}
\date{\today} 

\begin{abstract}
Poisson-Boltzmann theory is the cornerstone for soft matter electrostatics.
We provide novel exact analytical solutions to this non-linear mean-field 
approach, for the diffuse layer of ions in the vicinity of a planar or 
a cylindrical macroion.
While previously known solution are for homogeneously charged objects,
the cases worked out exhibit a modulated surface charge --or equivalently 
surface potential-- on the macroion (wall) surface. 
In addition to asymptotic features at large distances from the wall, 
attention is paid to the fate of the contact theorem, relating the contact 
density of ions to the local wall charge density. 
For salt-free systems (counterions only), we make use of results pertaining to 
the two-dimensional Liouville equation, supplemented by an inverse approach. 
When salt is present, we invoke the exact two-soliton solution to 
the 2D sinh-Gordon equation. 
This leads to inhomogeneous charge patterns, that are either localized or 
periodic in space. 
Without salt, the electrostatic signature of a charge pattern on the macroion 
fades exponentially with distance for a planar macroion, while it decays 
as an inverse power-law for a cylindrical macroion. 
With salt, our study is limited to the planar geometry, and reveals that 
pattern screening is exponential.
\end{abstract}

\maketitle

\renewcommand{\theequation}{1.\arabic{equation}}
\setcounter{equation}{0}

\section{Introduction} \label{Sec1}
Charges are omnipresent at the microscopic level in soft matter and biological 
systems \cite{Andelman06}. 
In a solvent like water, featuring efficient solvation and screening properties,
surface groups dissociate from large macromolecules (colloids), which results 
in mobile counterions in the vicinity of charged surfaces. 
While mobile ions are generically of both signs (both co- and counter-ions), 
it is possible to approach experimentally the limit of deionized  --or 
salt-free--  suspensions \cite{Palberg04}, where co-ions are absent.
This provides a convenient venue for theoretical investigations, that have 
studied thermal equilibrium both in the weak-coupling  
\cite{Attard88,Podgornik90,Netz00} and in the strong-coupling
\cite{Moreira00,Netz01,Moreira02a,Kanduc07,Jho08,Kanduc08,Samaj11,Samaj16,Samaj18}
regimes.

A pillar for the theoretical description of the structure of mobile ions in 
the vicinity of charged colloids, the so-called electric double-layer, 
is provided by the Poisson-Boltzmann theory (PB). 
It dates back to the pioneering works of Gouy \cite{Gouy10} and 
Chapman \cite{Chapman13} more than a century ago: it amounts to relating 
the local charge density appearing in Poisson equation to 
the Boltzmann weight of the mean electrostatic potential.
In doing so, one considers the mobile charged species as an inhomogeneous 
ideal gas, in a self-consistently determined (although external) electric 
field, see e.g. the reviews \cite{Attard96,Hansen00,Levin02,Messina09}.
Electrostatic and steric correlations are thereby neglected, an approach which 
requires to work in the weak coupling regime.
Such a mean-field approximation led to the DLVO  theory \cite{Verwey48}, that 
proved essential for rationalizing colloidal interactions.

Analytical solutions of electrostatic theories are useful, allowing 
to understand the combined effects for the different parameters, 
such as charge density, temperature, solvent or electrolyte type etc.
Screening properties in particular stand foremost, and will receive particular 
attention below.
Previously known explicit analytical exact solutions to the Poisson-Boltzmann 
theory are scarce unfortunately, essentially limited to 
\begin{itemize}
\item 
a single uniformly charged infinite plate with or without salt, 
two plates or a collection of equi-spaced parallel such plates in the 
salt-free case \cite{Andelman06,Attard96,Hansen00,Levin02,Messina09,Polat}.
Such a geometry is relevant for studying lamellar phases \cite{Dubois98}.
\item 
a cylindrical colloid, such as DNA, when bending and edge effects are 
neglected, leading to the infinite cylinder model. 
Exact results were obtained in the 1950s for a cylindrical concentric 
Wigner-Seitz cell without salt \cite{Alfrey}.
This solution appears as a restricted version of that for a partial 
differential equation first studied and solved by Liouville in the 1850s 
\cite{Liouville1853}. 
More recently, Tracy and Widom obtained a nontrivial exact solution 
for a single infinite straight and homogeneously charged line \cite{Tracy}. 
Exact but perturbative treatments were proposed to account for 
the finite extension of the charged cylinder \cite{TT06}, which in turn led
to an accurate description for the persistence length of semi flexible 
polymers \cite{ShenTrizac,Guilbaud}.
\end{itemize}
To the best of our knowledge, no exact result has been reported for 
heterogeneously charged macroions. 
It is our purpose here to put forward a number of such solutions, 
with or without salt, and to discuss the corresponding screening features. 
To this end, two techniques will be advocated: the two-dimensional (2D) 
Liouville artillery for salt-free systems, and the soliton method for 
solving the 2D sinh-Gordon equation. 
Since these approaches are two-dimensional in spirit, their translation to a 
three-dimensional (3D) problem necessarily leads to invariance along one 
Cartesian coordinate, see below.

A number of experimental ``anomalies'' --pertaining to particle flocculation, 
adhesion or deposition-- have been attributed to charge heterogeneities
or patterns \cite{Walz98}.
Early theoretical studies of systems with surface charge modulations were 
based on liquid-state approximations \cite{Chan80,Kjellander88b,Gonzalez01}.  
The combination of Monte-Carlo simulations with analytic perturbation 
techniques to charged-modulated surfaces in the strong coupling 
\cite{Moreira02b} and weak or intermediate coupling \cite{Lukatsky02a,Henle04} 
regimes indicates, for the studied forms of modulations, an increase of 
the mean counterion density close to the inhomogeneously charged surfaces, 
in comparison with that for the uniformly charged surfaces of 
the same averaged charge density.
The notable amplification of this counterion surface enhancement occurs at 
planar surfaces with disordered surface charge distributions \cite{Fleck05}.
For two parallel charge-modulated surfaces the enhancement of counterion
density near the surfaces means less charges at the midplane, and therefore
leads to a reduction of the pressure between the charged plates
\cite{Lukatsky02b,Khan05,rque10}.

Our interest will be twofold, with focus on both short distance and 
long-distance features. 
In the former category, relating the ionic density at contact
with the wall, to the surface charge, is of particular interest.
For the geometry of one uniformly charged planar wall, the contact theorem 
provides an exact and particularly simple answer
\cite{Henderson78,Henderson79,Carnie81,Wennerstrom82}, 
see the review \cite{Blum92}.
The generalization of the contact relation to curved wall boundaries
was the subject of a number of studies 
\cite{Blum94,Trizac97,Mallarino15,Malgaretti18}.
Here, we construct a PB generalization of the contact relation between 
the density profile at the wall and the inhomogeneous surface charge density, 
based on the fact that the total force acting on the wall must vanish 
in thermal equilibrium. 
All exact solutions fulfill this nonlocal contact relation, but interestingly,
some of the solutions provide a local relationship between the total particle 
number density at the wall surface and the inhomogeneous surface charge density.

Turning to long-distances properties, the decay of density profiles depends 
on the model under scrutiny. For charged plates, we will show that 
the influence of a charge pattern on the surface decays exponentially fast 
away from the plate, irrespective of the presence of salt. 
This applies in particular to the planar no-salt case, where the density 
profile goes to zero at large distances from the wall more slowly than 
an exponential, as an inverse power-law of type $1/x^2$ \cite{Andelman06}. 
This asymptotic behavior is universal in the sense that it does not depend
on the strength of the surface charge density.
For a cylindrical macroion, a charge pattern on the surface of 
the cylinder extends further than for plates, with a pattern screening of 
inverse power-law type, the exponent of which will be worked out.
 
The paper is organized as follows.
The general formulation of the models studied, together with their PB treatment
are given in Sec. \ref{Sec2}. 
The contact relation between the particle density and the surface charge 
density, known hitherto for uniformly charged plates, is generalized 
to modulated surface charges.
Based on the general solution of the 2D Liouville equation, exactly solvable 
cases for surface charge modulations with counterions only 
are generated in an inverse fashion in Sec. \ref{Sec3}.
The explicit results for the potential and particle density are analyzed
close, and far away from the charged interface. 
The results are relevant for both planar and cylindrical geometries.
Sec. \ref{Sec4} deals with models with added salt.  
The case of small charges is first worked out (Debye-H\"uckel perturbative 
treatment).
The exact non-perturbative 2-soliton solution of the nonlinear 2D sinh-Gordon
equation is then presented. 
Sec. \ref{Sec5} brings a short recapitulation of the most important results.

\renewcommand{\theequation}{2.\arabic{equation}}
\setcounter{equation}{0}

\section{General framework} \label{Sec2}
\subsection{Relevant boundary conditions} 
We are interested in the electrostatic potential created by a charged macroion 
in the 3D Euclidean space of points ${\bf r}=(x,y,z)$.
It is sufficient here to restrict our study to the exterior of the macroion, 
a region that we shall denote as $\Lambda$. 
The presence of the macroion materializes through the boundary conditions 
fulfilled by the potential. 
Classical point-like particles of (say elementary) charge $e$ can move 
in $\Lambda$.
They are immersed in a medium of dielectric constant $\varepsilon$. 
The wall surface carries a fixed surface charge density $\sigma e$,
that can be position dependent. 
The system is in thermal equilibrium at some inverse temperature
$\beta=1/(k_{\rm B}T)$. 

The interaction energy of two charges $q$ and $q'$ at the 
points ${\bf r}$ and ${\bf r}'$ in $\Lambda$ is given by
$q q'/(\varepsilon \vert {\bf r}-{\bf r'}\vert)$.
The Bjerrum length 
\begin{equation} \label{lB}
\ell_{\rm B} = \frac{\beta e^2}{\varepsilon}
\end{equation}
is the distance between two unit charges at which they interact with 
thermal energy $k_{\rm B}T$.
For a {\em uniform} surface charge density $\sigma e$, there exists another
relevant length scale. 
Since the potential energy of a unit charge at distance $z$ from such a
wall  is
$2\pi e^2\sigma z/\varepsilon$, this energy equals to $k_{\rm B}T$ at 
the so-called Gouy-Chapman length 
\begin{equation} \label{mu}
\mu = \frac{1}{2\pi\ell_{\rm B}\sigma} .
\end{equation}
The introduction of $\mu$  is {\it a priori} meaningful only for 
uniform surface charge densities. 

Let $\rho({\bf r})$ be the mean charge density of particles at point
${\bf r}\in\Lambda$. 
Denoting by $\psi({\bf r})$ the corresponding mean electrostatic
potential, the electric field is given by 
\begin{equation}
{\bf E} = - \nabla\psi.
\end{equation}
The electric field can be decomposed into its perpendicular and parallel 
components with respect to the wall surface (that may be curved, see below
the cylindrical geometry): 
${\bf E} = (E_{\perp},{\bf E}_{\parallel})$ where
\begin{equation}
E_{\perp} = - \frac{\partial\psi}{\partial x} , \qquad
{\bf E}_{\parallel} = - \left( \frac{\partial\psi}{\partial y} ,
\frac{\partial\psi}{\partial z} \right) . 
\end{equation}
Gauss's law demands that \cite{Jackson98} 
\begin{equation}
\nabla\cdot {\bf E} = \frac{4\pi}{\varepsilon} \rho
\end{equation}
and the mean potential therefore fulfills the Poisson equation 
\begin{equation} \label{Poisson}
\Delta \psi = - \frac{4\pi}{\varepsilon} \rho .
\end{equation}
The surface charge density $\sigma e$ is related to the normal derivative of 
$\psi$ at the wall as follows \cite{Jackson98}
\begin{equation} \label{BCorig1}
\frac{\partial\psi(x,y,z)}{\partial x}\Big\vert_{x=0} = 
- \frac{4\pi\sigma(y,z) e}{\varepsilon} . 
\end{equation} 
The overall system charge neutrality requires that the electric field 
vanishes at infinite distance from the wall:
\begin{equation} \label{BCorig2}
\lim_{x\to\infty} \frac{\partial\psi}{\partial x} \to 0 ,
\end{equation}
where $x$ is a proxy for the distance to the charged macroion. 
We thus consider here the infinite dilution limit, with a single, 
field creating, charged body.

\subsection{Poisson-Boltzmann theory}
We will address two distinct situations:
\begin{itemize}
\item
For counterions only systems, all mobile particles have 
the same charge, say $-e$.
Denoting by $n({\bf r})$ the particle number density at point 
${\bf r}\in\Lambda$, the charge density is simply given by 
$\rho({\bf r}) = -e n({\bf r})$.
Due to the requirement of  overall electroneutrality, the particle 
density must vanish at asymptotically large distances from the wall, i.e. 
\begin{equation} \label{BCdensity}
\lim_{x\to\infty} n({\bf r}) \to 0.
\end{equation}
In the standard mean-field approach, the particle density at a given point 
is proportional to the Boltzmann weight of the mean electrostatic potential 
at that point \cite{Andelman06}, 
\begin{equation}
n({\bf r}) = f_0 \exp\left[ \beta e \psi({\bf r}) \right] , 
\end{equation}
where $f_0$ is a normalization constant.
Introducing the reduced potential $\phi = \beta e\psi$, this mean-field 
assumption applied to (\ref{Poisson}) leads to the Poisson-Boltzmann (PB) 
equation
\begin{equation} \label{PB1}
\Delta \phi =  4\pi\ell_{\rm B}f_0 e^{\phi} , \qquad
n = f_0 e^{\phi} .
\end{equation}
Note a gauge freedom in shifting $\phi$ by a constant which only 
renormalizes $f_0$.
The boundary conditions (\ref{BCorig1}) and (\ref{BCorig2}) read 
\begin{equation} \label{BCreduced}
\frac{\partial\phi}{\partial x}\Big\vert_{x=0} = 
- 4\pi \ell_{\rm B} \sigma , \qquad 
\lim_{x\to\infty} \frac{\partial\phi}{\partial x} \to 0 . 
\end{equation} 
The asymptotic vanishing of the particle density (\ref{BCdensity}) means
that $\phi$ goes to $-\infty$ as $x\to\infty$. 
 
\item
For systems with salt, namely the symmetric two-component plasma,
we consider two kinds of mobile particles with charges $+e$ and $-e$.
Denoting the number densities of positively and negatively charged
species by $n_+({\bf r})$ and $n_-({\bf r})$, the total
particle number density is given by
\begin{equation}
n({\bf r}) = n_+({\bf r}) + n_-({\bf r})
\end{equation}
and the charge density reads as
\begin{equation}
\rho({\bf r}) = e \left[ n_+({\bf r}) - n_-({\bf r}) \right] .
\end{equation}
The system is electroneutral in the bulk region  $(x\to\infty)$, 
so that the bulk species densities must satisfy $n_+ = n_- = n/2$,
$n$ being the (prescribed) total bulk density of particles. 
Within the mean-field assumption for position-dependent species densities 
\begin{equation}
n_{\pm}({\bf r}) = n_{\pm} \exp[\mp \beta e\psi({\bf r})] ,
\end{equation}
$\psi({\bf r})$ must go to $0$ as $x\to\infty$. 
Defining the inverse Debye length $\kappa = \sqrt{4\pi\ell_{\rm B}n}$ and
with regard to the Poisson Eq. (\ref{Poisson}), the PB equation for 
the reduced potential $\phi = \beta e \psi$ reads as
\begin{equation} \label{PB2}
\Delta \phi({\bf r}) = \kappa^2 \sinh\phi({\bf r}) ,
\end{equation}
and we have
\begin{equation}
\phantom{aaa} n({\bf r}) = n \cosh\phi({\bf r}) , \quad
\rho({\bf r}) = - e n \sinh\phi({\bf r}) .
\end{equation}
The boundary condition (\ref{BCreduced}) for the reduced potential at 
the wall ($x=0$) remains unchanged, while the boundary conditions at a
symptotically large distances from the wall take the forms
\begin{equation}
\lim_{x\to\infty} \phi \to 0 , \qquad
\lim_{x\to\infty} \frac{\partial\phi}{\partial x} \to 0 . 
\end{equation}

\end{itemize}

\subsection{Generalization of the contact relation for a planar interface}
We aim at generalizing to the inhomogeneous case the contact relation between 
the particle and uniform surface charge densities. 
We restrict here to a planar wall, located at $x=0$.
We start with the definition of the pressure tensor in a charged medium
\cite{Landau84}
\begin{equation}
\tensor{\boldsymbol{\Pi}}({\bf r}) = \left[ k_{\rm B}T n({\bf r}) 
+ \frac{\varepsilon}{8\pi} {\bf E}^2({\bf r}) \right] \tensor{\boldsymbol{I}} 
- \frac{\varepsilon}{4\pi}
{\bf E}({\bf r}) \otimes {\bf E}({\bf r}) ,
\end{equation}
where $\tensor{\boldsymbol{I}}$ is the unity tensor.
The pressure tensor satisfies the mechanical equilibrium condition
\begin{equation}
\boldsymbol{\nabla} \cdot \tensor{\boldsymbol{\Pi}} = {\bf 0} . 
\end{equation}
A surface element $d{\bf S}$ of the wall at $x=0$, which is 
a vector perpendicular to the surface, is subject to the force
\begin{equation}
d{\bf F} = \tensor{\boldsymbol{\Pi}}\cdot d{\bf S} 
= \left[ k_{\rm B}T n + \frac{\varepsilon}{8\pi} 
\left( E_{\perp}^2 + {\bf E}_{\parallel}^2 \right) -
\frac{\varepsilon}{4\pi} E_{\perp}^2 \right] d{\bf S} , 
\end{equation}
where all quantities are dependent on the $(y,z)$-coordinates of the surface
element.
Now let us place a parallel planar wall with no surface charge at $x\to\infty$.
Since for neutral systems, ${\bf E}$ vanishes at $x\to\infty$, 
we have the force $d{\bf F}' = k_{\rm B}T n_{\rm bulk} d{\bf S}'$ where
the surface element on the oppositely oriented wall at $x\to\infty$ 
$d{\bf S}'= - d{\bf S}$ and $n_{\rm bulk}$ is the uniform bulk particle density.
The total (osmotic) pressure at point $(y,z)$ is thus given by
\begin{eqnarray} \label{pres}
P(y,z) & = & k_{\rm B}T \left[ n(0,y,z) - n_{\rm bulk} \right] \nonumber \\ & &
+ \frac{\varepsilon}{8\pi} \left[ {\bf E}_{\parallel}^2(0,y,z) 
- E_{\perp}^2(0,y,z) \right] . 
\end{eqnarray}

For a planar surface with uniform surface charge density $\sigma e$,
${\bf E}_{\parallel}=0$ and the quantities $n(0,y,z)=n(0)$ and
$E_{\perp}(0,y,z)=E_{\perp}(0)$ no longer depend on $y,z$-coordinates.
With regard to Eq. (\ref{BCorig1}) taken with $\sigma(y,z)=\sigma$, 
the requirement of the nullity of the pressure $P(y,z)=P$ leads to 
the standard contact relation
\begin{equation} \label{contact}
n(0) - n_{\rm bulk} = 2\pi\ell_{\rm B} \sigma^2 .
\end{equation}
We recall that $n_{\rm bulk}=0$ for charged walls with counterions only
and $n_{\rm bulk}=n$ when both co- and counterions are present (added salt).

In the inhomogeneous (patterned) case with position dependent surface charge
$\sigma(y,z) e$,  we have in general that ${\bf E}_{\parallel}\ne 0$.
The local pressure (\ref{pres}) can be expressed in terms of 
the reduced potential as follows
\begin{eqnarray} \label{P1}
\beta P(y,z) & = &  \left[ n(0,y,z) - n_{\rm bulk} \right]
- 2\pi\ell_{\rm B} \sigma^2(y,z) \nonumber \\ & &
+ \frac{1}{8\pi\ell_{\rm B}} \left\{
\left[ \frac{\partial\phi(0,y,z)}{\partial y} \right]^2
+ \left[ \frac{\partial\phi(0,y,z)}{\partial z} \right]^2 \right\} .
\nonumber \\ & &
\end{eqnarray}
It may be both positive or negative.
The mechanical condition for the plate equilibrium is that 
the {\em total} pressure exerted on the wall vanish, i.e.
\begin{equation} \label{C1}
\int_{-\infty}^{\infty} dy \int_{-\infty}^{\infty} dz  P(y,z) = 0 .
\end{equation}
Appendix \ref{app:Rederivation} offers a rederivation of Eqs. \eqref{P1} 
and \eqref{C1} directly from the PB equation.

In the case of a surface charge density varying only along one direction,
i.e. $\sigma(y,z)=\sigma(y)$, $\phi(x,y,z)=\phi(x,y)$ and $n(x,y,z)=n(x,y)$, 
one has the simplified expression for the pressure
\begin{equation} \label{P2}
\beta P(y) =  \left[ n(0,y) - n_{\rm bulk} \right]
- 2\pi\ell_{\rm B} \sigma^2(y) + \frac{1}{8\pi\ell_{\rm B}} 
\left[ \frac{\partial\phi(0,y)}{\partial y} \right]^2 
\end{equation}
and the constraint
\begin{equation} \label{C2}
\int_{-\infty}^{\infty} dy \beta P(y) = 0 .
\end{equation}
If the system is periodic along the $y$-axis with period ${\cal P}$, 
it is sufficient to integrate over this period, say
\begin{equation} \label{C3}
\int_0^{\cal P} dy \beta P(y) = 0 .
\end{equation}
The validity of the pressure constraint will be verified for every exactly 
solvable planar model.
For the uniformly charged wall $\sigma(y)=\sigma$ with ${\bf E}_{\parallel}=0$
the local relation $\beta P = 0$ applies.

Due to the positivity of $[\partial_y\phi(0,y)]^2$ in (\ref{P2})
the following inequality holds
\begin{equation} \label{ineq}
\int_{-\infty}^{\infty} dy \left[ n(0,y) - n_{\rm bulk} \right]
\leq 2\pi\ell_{\rm B} \int_{-\infty}^{\infty} dy \sigma^2(y) ,
\end{equation}
where the equality applies exclusively for the uniform case with
$\partial_y\phi(0,y) = 0$. 
Defining the average over the whole plate by brackets, 
inequality (\ref{ineq}) can be rewritten as
\begin{equation} \label{ineq1}
\frac{\langle n(0)\rangle - n_{\rm bulk}}{
2\pi\ell_{\rm B} \langle \sigma^2 \rangle} \leq 1 .
\end{equation}
In particular, if in an inhomogeneous model the relation between 
the contact particle density and the surface charge density is of local type
\begin{equation} \label{localrel}
n(0,y) - n_{\rm bulk} \propto 2\pi\ell_{\rm B} \sigma^2(y) ,
\end{equation}
the prefactor must be less than 1.
Local relations of type (\ref{localrel}) are rare, but they exist as
we shall see later.

Interestingly, relation \eqref{ineq1} may superficially seem at variance 
with the phenomenon of increased counterion condensation near surfaces, 
reported in \cite{Lukatsky02a,Henle04,Fleck05} for systems with counterions 
only ($n_{\rm bulk}=0$). 
This enhancement effect translates into
\begin{equation} \label{Lukatsky}
\frac{\langle n(0)\rangle}{2\pi\ell_{\rm B} \langle\sigma\rangle^2} > 1 .
\end{equation}
Such an inequality may be compatible with \eqref{ineq1}. 
In such a case, we have 
\begin{equation}
2 \pi \ell_{\rm B}  \langle\sigma\rangle^2  \,<\,   
\langle n(0)\rangle \, < \, 2 \pi \ell_{\rm B}  \langle \sigma^2 \rangle .
\end{equation}
While the upper bound is guaranteed, a pending question is thus whether 
on general grounds, 
\begin{equation} \label{Lukatsky2}
2\pi\ell_{\rm B} \langle\sigma\rangle^2     \,  \mathop{<}^?    \, 
\langle n(0)\rangle.
\end{equation}

\renewcommand{\theequation}{3.\arabic{equation}}
\setcounter{equation}{0}

\section{Salt-free systems (counterions only)} \label{Sec3}

\subsection{Uniform planar surface charge density} 
We first recapitulate the case of a uniform plate charge density $\sigma e$,
see e.g. review \cite{Andelman06}.  
The electrostatic potential and particle density then depend only on 
the $x$-coordinate.
Introducing its dimensionless counterpart
\begin{equation}
\widetilde{x} = \sqrt{2\pi\ell_{\rm B}f_0} \, x ,
\end{equation}
the PB equation (\ref{PB1}) can be written as
\begin{equation} \label{PBuniform}
\frac{d^2\phi}{d\widetilde{x}^2} = 2 e^{\phi}
\end{equation}
and the boundary condition (\ref{BCreduced}) at $\widetilde{x}=0$ 
takes the form 
\begin{equation} \label{BCuniform}
\frac{\partial\phi}{\partial \widetilde{x}}\Big\vert_{\widetilde{x}=0} = 
- \frac{4\pi \ell_{\rm B} \sigma}{\sqrt{2\pi\ell_{\rm B}f_0}} .
\end{equation} 
Multiplying the PB equation (\ref{PBuniform}) by $d\phi/d\widetilde{x}$
leads to
\begin{equation}
\frac{1}{2} \left( \frac{d\phi}{d\widetilde{x}} \right)^2 - 
2 e^{\phi}  = {\rm cst.}
\end{equation}  
The constant on the rhs of this equation vanishes due to 
the boundary conditions at $\widetilde{x}\to\infty$.
Setting the gauge $\phi(0)=0$, the  solution reads
\begin{equation} \label{phiLiou}
\phi = - 2 \ln(1+\widetilde{x}) .
\end{equation}
The normalization constant $f_0$ is fixed by the boundary condition 
(\ref{BCuniform}) to
\begin{equation}
f_0 = 2\pi\ell_{\rm B} \sigma^2 .
\end{equation}
Thus, $\widetilde{x}=x/\mu$ where $\mu$ is the Gouy-Chapman length 
defined in Eq. (\ref{mu}).
 
The particle number density has the form
\begin{equation} \label{nuniform}
n = 2\pi\ell_{\rm B}\sigma^2 \frac{1}{(1+\widetilde{x})^2} 
\mathop{\sim}_{x\to\infty} \frac{1}{2\pi\ell_{\rm B}} \frac{1}{x^2} .
\end{equation}
The long-ranged asymptotic decay is universal as it does not depend on 
the surface charge density $\sigma e$. It is power-law-like
as a result of poor screening (counterions only, no salt). 
The contact value of the number density $n(0) = 2\pi\ell_{\rm B}\sigma^2$
is in agreement with the contact theorem (\ref{contact}).

\subsection{General solution of the 2D Liouville equation}
Let us now consider a modulation of the surface charge density, say along
the $y$-axis.
The electrostatic potential and the particle density depend on coordinates 
$x$ and $y$.
Let us fix the normalization constant as follows
\begin{equation}
f_0 = \frac{1}{2\pi\ell_{\rm B}^3} .
\end{equation}
Introducing the dimensionless coordinates
\begin{equation}
\widetilde{x} = \frac{x}{\ell_{\rm B}} , \qquad
\widetilde{y} = \frac{y}{\ell_{\rm B}} , 
\end{equation}
the PB equation (\ref{PB1}) is written as
\begin{equation} \label{PBmodulated}
\frac{\partial^2\phi}{\partial \widetilde{x}^2} + 
\frac{\partial^2\phi}{\partial \widetilde{y}^2} = 2 e^{\phi} . 
\end{equation}  
The boundary condition (\ref{BCreduced}) at $\widetilde{x}=0$ has the form 
\begin{equation} \label{BCmodulated}
\frac{\partial\phi}{\partial\widetilde{x}}\Big\vert_{\widetilde{x}=0} = 
- 4\pi \ell_{\rm B}^2 \sigma(\widetilde{y})
\end{equation} 
and the particle density is expressible as
\begin{equation} \label{nmodulated}
n = \frac{1}{2\pi\ell_{\rm B}^3} e^{\phi} .  
\end{equation}
This relation will remain true
in the remainder, for all solutions worked out.

Eq. (\ref{PBmodulated}), known in the mathematical literature as the 2D 
Liouville partial differential equation \cite{Liouville1853}, 
is of elliptic type and has a number of applications in physics, 
in soft matter but also beyond, see e.g. \cite{Chavanis} for a study of 
the dynamics of point vortices.
Various partial exact solutions of this equation have been found in the past,
see e.g. \cite{Picard1898,Clarkson89,Bhutani94,Popov93}.
The most general {\em real} solution of the 2D Liouville equation  
has been found by Crowdy \cite{Crowdy97}.
In terms of the complex variables
\begin{equation} \label{cv}
z = \widetilde{x} + {\rm i} \, \widetilde{y} , \qquad
\bar{z} = \widetilde{x} - {\rm i} \, \widetilde{y} ,
\end{equation}
the general solution takes the form 
\begin{eqnarray}
\phi(\widetilde{x},\widetilde{y}) & = & 
- 2 \ln \left[ c_1 Y_1(z) \bar{Y}_1(\bar{z}) +
c_2 Y_2(z) \bar{Y}_2(\bar{z}) \right. \nonumber \\ & & \left.
+ c_3 Y_1(z) \bar{Y}_2(\bar{z}) + \bar{c}_3 \bar{Y}_1(\bar{z}) Y_2(z) \right] 
\nonumber \\ & & + \ln \left[ W(z) \bar{W}(\bar{z}) \right] . \label{gensol}
\end{eqnarray}
Here, $Y_1(z)$ and $Y_2(z)$ are two arbitrary but independent analytic
functions  with the nonzero Wronskian 
\begin{equation} \label{Wronskian}
W(z)\equiv Y_1(z) Y'_2(z) - Y'_1(z) Y_2(z) .
\end{equation}
$c_1$, $c_2$ are real constants and $c_3$ a complex constant, such that 
the constraint
\begin{equation} \label{constraint}
\vert c_3 \vert^2 - c_1 c_2 = \frac{1}{4} 
\end{equation}
is satisfied.
The conjugate function $\bar{f}(\bar{z})$ is defined by 
$\bar{f}(\bar{z}) = \overline{f(z)}$.

We shall use the above general solution in an inverse way, namely
generating from the electrostatic potential (which is a regular solution 
of the generic 2D Liouville equation) the corresponding surface charge density. 
There are two strong limitations on acceptable solutions.
Firstly, we are interested only in regular solutions which do not exhibit 
an unphysical singularity (divergence) at any point of the available space 
$\Lambda$.
This means that expressions under logarithms must always be positive.
Secondly, many exact solutions correspond to non-neutral systems, 
and have been consequently discarded.

The simplest solution is given by the functions 
\begin{equation}
Y_1(z) = z , \qquad Y_2(z) = 1 ,
\end{equation}
with the Wronskian $W(z) = -1$.
Writing $c_3=\alpha + i \beta$, the solution for the potential reads
\begin{equation} \label{phisimplest}
\phi = - 2 \ln \left[ c_1 (\widetilde{x}^2+\widetilde{y}^2) 
+ c_2 + 2\alpha \widetilde{x} - 2 \beta \widetilde{y} \right] ,
\end{equation}
where the parameters are constrained by
\begin{equation} \label{constraintsimplest}
\alpha^2 + \beta^2 - c_1 c_2 = \frac{1}{4} .
\end{equation}
For the choice $c_1=0 , c_2=c , \alpha=1/2 , \beta=0 $,
one has the $\widetilde{y}$-independent potential
\begin{equation}
\phi = - 2 \ln \left( c + \widetilde{x} \right) .
\label{eq:solbasicplanar}
\end{equation}
$c$ is related to the constant surface charge density $\sigma$ via 
the boundary condition (\ref{BCmodulated}) as follows
\begin{equation}
c = \frac{1}{2\pi\ell_{\rm B}^2\sigma} = \frac{\mu}{\ell_{\rm B}} .
\end{equation} 
With regard to (\ref{nmodulated}), the profile of particle density
\begin{equation}
n = \frac{1}{2\pi\ell_{\rm B}^3} \frac{1}{(c+\widetilde{x})^2}
= 2\pi\ell_{\rm B}\sigma^2 \frac{1}{(1+x/\mu)^2} 
\end{equation}
coincides with the previous one (\ref{nuniform}).
Up to the different choice of units, we recover the plain solution 
\eqref{phiLiou}.
As we shall see, other choices of the building blocks $Y_1$ and $Y_2$ 
yield more interesting results. Before we proceed along these lines,
we present a useful mapping between planar and cylindrical geometry,
that allows a one to one correspondence. 

\subsection{Towards heterogeneous charge distributions} \label{l}
If a given solution $\phi(\widetilde x, \widetilde y)$ is known for 
Eq. \eqref{PBmodulated}, it is straightforward to realize that 
\begin{equation}
\phi_{\text{cyl}}(\widetilde r,\varphi) = \phi(\widetilde x,\widetilde y) 
-2\ln \widetilde r \label{eq:mapping}
\end{equation}
with $\widetilde r = e^{\widetilde x}$ and $\varphi=\widetilde{y}$ 
also obeys the PB equation
\begin{equation}
\nabla^2 \phi_{\text{cyl}} \, = \, 2  \, e^{\phi_{\text{cyl}}}
\end{equation}
in cylindrical coordinates, where the Laplacian takes the form 
\begin{equation}
\nabla^2  \,= \, \frac{1}{\widetilde r} \frac{\partial}{\partial \widetilde r} 
\left( \widetilde r \frac{\partial}{\partial \widetilde r} \right) \,+\, 
\frac{1}{\widetilde r^2}\frac{\partial^2}{\partial \varphi^2} .
\end{equation}
This mapping has been invoked in Ref. \cite{BuOr06} and can already be found 
in the pioneering work of Fuoss {\it et al.\/} on charged rods in Wigner-Seitz cells
\cite{Alfrey}.
It yields a one to one correspondence between a solution in planar geometry 
(expressed with Cartesian coordinates), 
and another one in cylindrical geometry.
It is interesting to note that while the planar solution associated to some 
$\phi(\widetilde x, \widetilde y)$ is electrically neutral 
(meaning that Eq. \eqref{BCorig2} holds), the cylindrical partner solution 
$\phi_{\text{cyl}}(\widetilde r,\varphi)$ is not: we indeed get that 
\begin{equation}
\widetilde r \phi'_{\text{cyl}}(\widetilde r,\varphi) \,=\,   
\phi'(\ln\widetilde r,\widetilde{y}) - 2 ,
\end{equation}
where the prime denotes derivative with respect to the first argument of 
the functions considered.
From Gauss theorem, this implies that the electric charge enclosed by 
a cylinder of divergent radius tends to $2/\ell_B$ per unit height of 
the cylinder. 
This is nothing but a manifestation of the well documented Manning evaporation 
phenomenon \cite{Oosawa68,Manning69,BuOr06,Juan}:
the logarithmic potential created by a bare charged cylinder is not 
sufficiently strong for confining all neutralizing counterions. 
Some counterions evaporate ``to infinity'', so that the integrated charge 
(cylinder plus localized counterions seen from the distance) amounts to 
the afore mentioned effective lineic value.

The planar $\leftrightarrow$ cylindrical mapping is useful to generate solutions
in both geometries, from a known solution. In doing so, we circumvent 
having to find the appropriate couple of generating functions $Y_1$ and $Y_2$. 
For instance, the cylindrical counterpart of the basic planar solution 
\eqref{eq:solbasicplanar}
reads 
\begin{equation}
 \phi_{\text{cyl}}  \, = \, -2 \ln \widetilde r   
- 2 \ln \left( c + \ln \widetilde r \right) . \label{eq:cylbasic}
\end{equation}
It turns that this solution is associated to the choice
\begin{equation}
Y_1(z)=\ln(z+a)+c , \qquad  Y_2(z)=1, 
\end{equation}
with real parameters $a, c$ constrained by
\begin{equation}
a,c\in R , \quad a,c > 0 , \quad \ln a + c > 0 
\end{equation} 
and the Wronskian $W(z)=-1/(z+a)$.
The constraint (\ref{constraint}) is met by setting
\begin{equation}
c_1 = c_2 = 0 , \qquad c_3=\frac{1}{2}.
\end{equation}
The associated general solution (\ref{gensol}) yields
\begin{eqnarray}
\phi & = & - 2 \ln \left\{ c + \frac{1}{2}
\ln\left[ (\widetilde{x}+a)^2+\widetilde{y}^2 \right] \right\} \nonumber \\
& & - \ln\left[ (\widetilde{x}+a)^2+\widetilde{y}^2 \right] .
\label{eq:backtocart}
\end{eqnarray}
which is nothing but \eqref{eq:cylbasic} expressed in Cartesian coordinates,
where $\widetilde r^2 = (\widetilde{x}+a)^2+\widetilde{y}^2$.
The corresponding surface charge density computed on the plate at $x=0$ reads
\begin{equation} \label{sigmanext}
\sigma = \frac{1}{2\pi\ell_{\rm B}^2} \frac{a}{a^2+\widetilde{y}^2}
\left[ 1 + \frac{1}{c+\frac{1}{2}\ln(a^2+\widetilde{y}^2)} \right] .
\end{equation}
It is, expectedly, localized in the vicinity of $\widetilde y=0$.
It implies a density profile of the form
\begin{equation} 
n = \frac{1}{2\pi\ell_{\rm B}^3} \frac{1}{(\widetilde{x}+a)^2+\widetilde{y}^2}
\frac{1}{\left\{ c + \frac{1}{2}\ln\left[ (\widetilde{x}+a)^2+\widetilde{y}^2
\right] \right\}^2} 
\end{equation}
which decays at asymptotically large distances from the wall as 
\begin{equation} \label{nnextasymp}
n(x,y) \mathop{\sim}_{x\to\infty} 
\frac{1}{2\pi\ell_{\rm B}} \frac{1}{x^2 (\ln x)^2} .
\end{equation}
This density falloff is faster than the one $\propto 1/x^2$ for the uniformly 
charged wall. 
This stems from the fact that the surface charge on the plate at $x=0$ is no 
longer uniform, but $y$-dependent and localized, with thus less strength 
to localize the counterions.
The decay is universal, independent of the surface charge 
characteristics $a$ and $c$.

To summarize, starting from the basic planar solution 
\eqref{eq:solbasicplanar}, we invoked the general mapping \eqref{eq:mapping} 
to generate a simple solution of PB equation (potential created by 
a uniformly charged cylinder), which in terms we have re-expressed in 
Cartesian coordinates to arrive at Eq. \eqref{eq:backtocart}
to generate the non-trivial solution for a non-uniformly charged plate with 
surface charge density \eqref{sigmanext}.
Yet, the latter planar solution is of limited interest and in some sense 
artificial, since it expresses in a set of coordinates (here Cartesian), 
the potential created by a body featuring a cylindrical symmetry. 
In the remainder, we will limit such considerations to Appendix 
\ref{app:nonneutral}, and consider solutions to the PB equation 
that are truly non-trivial. 
A key question has to do with the screening effects pertaining to 
an inhomogeneous periodic surface charge density.

\subsection{Periodic modulations of the surface charge} \label{p}

\subsubsection{Planar formulation} \label{ssec:planar}
To generate periodically changing surface charge densities, we propose
the following functions
\begin{equation} \label{ansatz}
Y_1(z) = z + a e^{-bz} , \qquad Y_2(z) = 1
\end{equation}
and the coefficients
\begin{equation} \label{coef}
c_1 = 0 , \quad c_2 = c , \quad c_3 = \frac{1}{2}
\end{equation}
which fulfill the constraint (\ref{constraint}).
At this stage, the free parameters $a, b, c$ are supposed to be positive real 
numbers:
\begin{equation}
a, b, c \in R , \qquad a, b, c > 0 .
\end{equation} 
The resulting potential has the form
\begin{eqnarray}
\phi & = & - 2 \ln \left[ c + \widetilde{x} + a e^{-b\widetilde{x}} 
\cos(b\widetilde{y}) \right] \nonumber \\ & &
+ \ln\left[ 1 - 2 a b e^{-b\widetilde{x}} \cos(b\widetilde{y}) 
+ (a b)^2 e^{-2b\widetilde{x}} \right] . 
\label{pot}
\end{eqnarray}
The regularity of $\phi$ in the domain $\Lambda$ requires that
\begin{equation} \label{inequal}
c > a , \qquad a b < 1 .
\end{equation}

The surface charge pattern generated from the potential (\ref{pot}),
\begin{equation} \label{sigmap}
\sigma = \frac{1}{4\pi\ell_{\rm B}^2} \Big\{
\frac{2(1+bc)}{c+a\cos(b\widetilde{y})} 
- \frac{b\left[1-(a b)^2 \right]}{1-2ab\cos(b\widetilde{y}) + (ab)^2} 
- b \Big\}
\end{equation}
is a periodic function of $\widetilde{y}$ with period $2\pi/b$.
Depending on the parameters $a$, $b$ and $c$, it can be both positive
and negative, but its mean value (naturally calculated over the period)
\begin{eqnarray}
\langle \sigma \rangle & \equiv & \frac{b}{2\pi} 
\int_0^{2\pi/b} d\widetilde{y} \sigma(\widetilde{y}) \nonumber \\ 
& = & \frac{b}{2\pi\ell_{\rm B}^2} \left[ \left( 1 + \frac{1}{bc} \right) 
\frac{1}{\sqrt{1-\left( \frac{a}{c}\right)^2}} - 1 \right]
\end{eqnarray}
is always positive as it should be in order to have negatively charged
particles in half-space $\Lambda$.

The particle density profile reads
\begin{equation}
n = \frac{1}{2\pi\ell_{\rm B}^3} \frac{1 
- 2 a b e^{-b\widetilde{x}} \cos(b\widetilde{y}) 
+ (a b)^2 e^{-2b\widetilde{x}}}{\left[ c + \widetilde{x} + a e^{-b\widetilde{x}} 
\cos(b\widetilde{y}) \right]^2} . \label{eq:ndecay}
\end{equation}
The exponential terms are negligible at large distances from the wall
and we recover the universal asymptotic decay of the particle density 
\begin{equation} 
n \mathop{\sim}_{x\to\infty} \frac{1}{2\pi\ell_{\rm B}} \frac{1}{x^2} ,
\end{equation}
exactly the same as in the uniform case (\ref{nuniform}).
This means that the periodic variation of the particle density due to
the surface charge density is suppressed exponentially fast,
in spite of poor screening properties of the charged 
system which normally imply a slow decay of statistical quantities.
The simultaneous appearance of short-ranged and long-ranged decays 
in the particle density profile is an interesting and unexpected feature
of the inhomogeneously charged surfaces. Eqs. \eqref{pot} and 
\eqref{eq:ndecay} reveal that ``memory'' of corrugation of the surface
is exponentially suppressed with distance $x$ from the plate; 
the corresponding decay length is $b^{-1}$, thus set by the periodicity of 
the charge ``pattern'' at $x=0$.
On the other hand, the mean density decays as a power-law.

The mean value of the particle density at the wall reads as
\begin{eqnarray}
\langle n(0) \rangle & \equiv & \frac{b}{2\pi} 
\int_0^{2\pi/b} d\widetilde{y} n(0,\widetilde{y}) \nonumber \\ 
& = & \frac{1}{2\pi\ell_{\rm B}^3 c^2} 
\frac{1 + (a b)^2 + 2 \frac{a^2b}{c}}{\left[ 
1 -\left(\frac{a}{c}\right)^2 \right]^{3/2}} .
\end{eqnarray}
Introducing the new parameters
\begin{equation} \label{parameters}
a b \equiv \alpha \in (0,1) , \qquad
\frac{a}{c} \equiv \beta \in (0,1) ,
\end{equation}
we derive for the ratio of interest (\ref{Lukatsky}) 
\begin{equation}
\frac{\langle n(0)\rangle}{2\pi\ell_{\rm B} \langle\sigma\rangle^2}
= \frac{1+\alpha^2+2\alpha\beta}{\sqrt{1-\beta^2}} 
\frac{\beta^2}{\left( \alpha+\beta-\alpha\sqrt{1-\beta^2}\right)^2} .
\end{equation}
The expression on the rhs of this equation is always bigger than or equal
to 1 within the definition regions (\ref{parameters}) of the parameters $\alpha$
and $\beta$; the unity value is obtained in the limit $\alpha,\beta\to 0$.
This confirms the previous findings about the enhancement of the counterion
density close to the wall \cite{Lukatsky02a,Henle04,Fleck05}.
Recalling the general contact inequality (\ref{ineq1}), the mean contact 
particle density of the present model has clear lower and upper bounds:
\begin{equation} \label{lowerupper}
2\pi\ell_{\rm B} \langle\sigma\rangle^2 \le \langle n(0)\rangle 
\le 2\pi\ell_{\rm B} \langle\sigma^2\rangle .
\end{equation}

Conservation laws take a simple form in periodic systems as the integrals
over the whole $\widetilde{y}$-axis are substituted by the ones over
one period.
The system's overall electroneutrality requires that
\begin{equation}
\ell_{\rm B} \int_0^{2\pi/b} d\widetilde{y} \left[ e \sigma(\widetilde{y})
+ \ell_{\rm B} \int_0^{\infty}d\widetilde{x} (-e) 
n(\widetilde{x},\widetilde{y}) \right] = 0
\end{equation}
and this equality was checked to be true.
The contact Eqs. (\ref{P2}) and (\ref{C3}) for the pressure also hold.

A straightforward generalization of the ansatz (\ref{ansatz}) is
\begin{equation}
Y_1(z) = z + \sum_n a_n e^{-b_n z} , \qquad Y_2(z) = 1 ,
\label{eq:ansatz}
\end{equation}
where $\{ a_n\}$ and $\{ b_n\}$ are any sets of positive real numbers, 
parameters $\{ b_n\}$ are distinct.
The constants $c_1$, $c_2$ and $c_3$ are chosen as in (\ref{coef}).
Under the constraints $c>\sum_n a_n$ and $\sum_n a_n b_n< 1$,
the resulting electrostatic potential and density profile contain 
superpositions of $\cos$-functions with different periods along
the $\widetilde{y}$-axis. We come back to these solutions below 
when discussing ``mode mixing''.

\subsubsection{Cylindrical formulation}
To transpose the previous periodically corrugated charge pattern on a plane,
to a periodic pattern on a cylinder, we take advantage of 
the mapping $z\to \ln z$ defined in section \ref{l}, 
or equivalently take 
\begin{equation}
 Y_1(z)  = \ln z + a z^{-b} ,  Y_2(z) = 1 ,  c_1=0, c_2=c, c_3=\frac{1}{2},
\end{equation}
with $b=1,2\ldots$. 
We then obtain 
\begin{eqnarray}
\phi(\widetilde r,\varphi)  &= & -2 \ln\widetilde r \nonumber \\
& & - 2 \ln \left[ c + \ln \widetilde r + \frac{a}{\widetilde r^b} 
\cos(b\varphi) \right] \nonumber \\ & &
+ \ln\left[ 1 - \frac{2 a b}{ \widetilde r^{b}} \cos(b\varphi) 
+ \frac{(a b)^2}{ \widetilde r^{2b} } \right] . \label{eq:potcyl}
\end{eqnarray}
The corresponding surface charge on a cylinder can be computed (with arbitrary 
radius as long as the quantities under $\ln$ are positive, which precludes 
too small radii). 
It is not our purpose to detail the precise result, since it is sufficient to 
note that it corresponds to a periodic pattern (or mode), with period $2\pi/b$. 
Eq. \eqref{eq:potcyl} indicates that this charge pattern has a signature 
in the potential, that decays as the inverse power-law $\widetilde r^{-b}$. 
Unlike in the periodic planar case where corrugation screening is exponential, 
the pattern is here screened algebraically. 
This ``duality'' appears generic ; it can be viewed as subsumed in the
planar to cylindrical mapping of section \ref{l}, and stems from the 
correspondence $\widetilde x \leftrightarrow \ln \widetilde r$.

\begin{figure}
\begin{center}
\includegraphics[width=0.235\textwidth,clip]{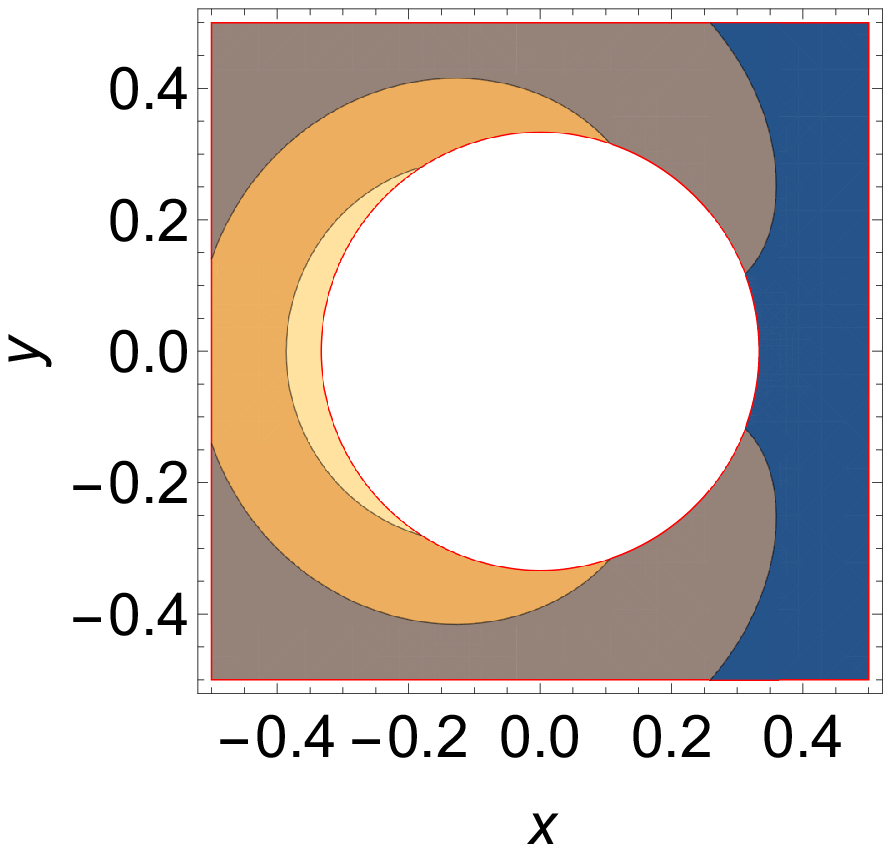}
\includegraphics[width=0.22\textwidth,clip]{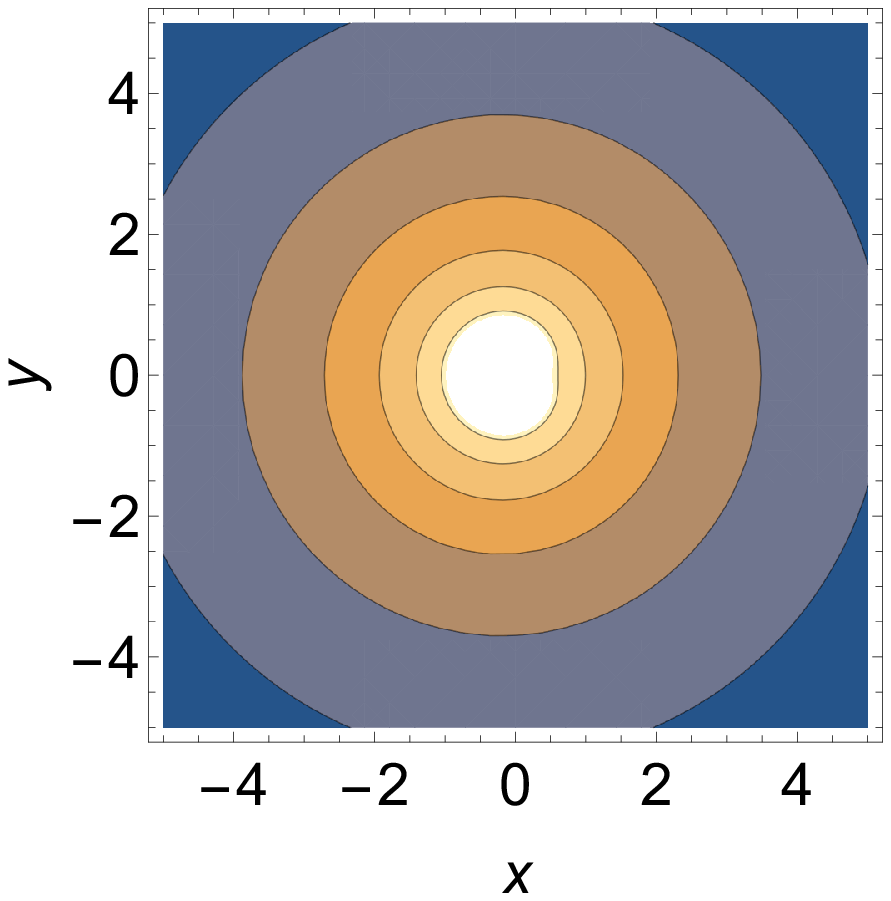}
\caption{Iso-potential lines associated to \eqref{eq:potcyl}, for $b=1$,  
$c=2$ and $a=0.2$.
The left panel shows the charged cylinder creating the field (central circle),
together with a zoom of the $\phi$ map. 
The right panel shows the iso-potentials on a larger scale. 
Long-distance isotropy becomes apparent.}
\label{fig:contour}
\end{center}
\end{figure}

For completeness Fig. \ref{fig:contour} shows the contour plot of $\phi(x,y)$
for the lowest order mode ($b=1$). 
It appears that the iso-potential lines become more and more isotropic, 
moving away from the charged cylinder shown in red on the left hand panel.

Here also, an ansatz of the form \eqref{eq:ansatz} yields a family of 
new solutions.
We now discuss the most salient feature of these generalized solutions.

\subsubsection{Mode mixing}
In the light of our previous cylindrical/planar mapping remark, 
we will discuss here the planar case only, keeping in mind 
the correspondence between exponential pattern screening for planes,
and algebraic pattern screening for cylinders (with an exponent related 
to the period of the periodic charge/potential pattern on the surface of 
the cylinder).

We generalize Eq. \eqref{ansatz} into 
\begin{equation} \label{ansatz:mixing}
Y_1(z) \, = \, z + a_1 e^{-b_1 z} + a_2 e^{-b_2 z} , \qquad Y_2(z) = 1
\end{equation}
with $b_1,b_2>0$.
This lead to the potential $\phi(\widetilde x,\widetilde y)$ given by
\begin{eqnarray}
\phi & = & - 2 \ln \left[ c + \widetilde{x} + a_1 e^{-b_1\widetilde{x}} 
\cos(b_1\widetilde{y}) + a_2 e^{-b_2\widetilde{x}} \cos(b_2\widetilde{y})
\right] \nonumber \\ & & + 
 \ln\left\{ 1 - 2 a_1 b_1 e^{-b_1\widetilde{x}} \cos(b_1\widetilde{y}) 
- 2 a_2 b_2 e^{-b_2\widetilde{x}} \cos(b_2\widetilde{y})   \right.
\nonumber \\ & & +   (a_1 b_1)^2 e^{-2b_1\widetilde{x}} +   
(a_2 b_2)^2 e^{-2b_2\widetilde{x}} \nonumber \\ & & + 
2 \left. a_1 b_1 a_2 b_2 \,  e^{-(b_1+b_2) \widetilde{x}} \,
\cos[(b_1-b_2)\widetilde y] \right\}  \label{potmixing}
\end{eqnarray}
and a more complex ``two-mode'' pattern on the plane at $x=0$ than 
in section \ref{ssec:planar}.
We see that considering two modes with periods 
(in coordinate $\tilde{y}$) $2\pi/b_1$ and $2\pi/b_2$ in the function ansatz
(\ref{ansatz:mixing}) implies in the potential solution (\ref{potmixing})
the corresponding decay lengths (in coordinate $\tilde{x}$) $1/b_1$ and 
$1/b_2$, respectively.
Yet, the contrubution dominating at long distances is not necessarily the one having 
the largest period since there is a contribution with period 
$2\pi/\vert b_1-b_2\vert$, which may possibly be the largest one, 
with a small decay length $1/(b_1+b_2)$.
From these results, we can surmise that the Fourier transform of a given
charge pattern on the plate will not allow to identify the long-distance 
electrostatic signature of the plate, by searching for the mode with 
smallest wave number.
As outlined above, these results immediately transpose to the cylindrical 
geometry, upon changing terms like $e^{-(b_i) \widetilde{x}}$ into 
$(\widetilde r)^{-b_i}$. 


\subsection{A perturbative solution of the Liouville equation} \label{ppp}
We next propose a perturbative treatment of the 2D Liouville equation around 
the full solution of the uniform surface charge density by considering its 
periodic modulations with infinitesimally small amplitudes. 
This will confirm the conclusions of the previous sections. 

Let us add to the uniform potential solution (\ref{phiLiou}) an infinitesimal 
perturbation $\epsilon f(\widetilde{x},\widetilde{y})$ with $\epsilon\ll 1$:
\begin{equation}
\phi(\widetilde{x},\widetilde{y}) = - 2 \ln(c+\widetilde{x}) 
+ \epsilon f(\widetilde{x},\widetilde{y}) .
\end{equation}
The parameter $c>0$ will be related to the surface charge density subsequently.
Inserting this ansatz into the 2D Poisson-Boltzmann/Liouville equation 
(\ref{PBmodulated}) and expanding all functions up to terms linear in 
the small parameter $\epsilon$, the function $f(x,y)$ must obey 
\begin{equation} \label{feqLiou}
\frac{\partial^2 f}{\partial\widetilde{x}^2} + 
\frac{\partial^2 f}{\partial\widetilde{y}^2} = 
\frac{2}{(c+\widetilde{x})^2} f  .
\end{equation}

Using separation of variables 
\begin{equation}
f(\widetilde{x},\widetilde{y}) = \varphi(\widetilde{x}) \psi(\widetilde{y})
\end{equation}
the functions $\varphi$ and $\psi$ fulfill the second-degree ordinary 
differential equation
\begin{equation} \label{twodeLiou}
\frac{1}{\varphi} \frac{d^2\varphi}{d \widetilde{x}^2} = 
b^2 + \frac{2}{(c+\widetilde{x})^2} , \qquad
\frac{1}{\psi} \frac{d^2\psi}{d \widetilde{y}^2} = - b^2
\end{equation}
with $b$ a free positive real number.
The solution for $\psi$ is 
\begin{equation}
\psi = \cos \left( b \widetilde{y} \right) , 
\end{equation}
where the prefactor is set to unity for simplicity.
The solution for $\varphi$ reads 
\begin{equation}
\varphi = e^{-b\widetilde{x}} \left( b + \frac{1}{c+\widetilde{x}} \right) .
\end{equation}

The total potential
\begin{equation} \label{phifinalLiou}
\phi = - 2 \ln(c+\widetilde{x}) + \epsilon e^{-b\widetilde{x}} 
\left( b + \frac{1}{c+\widetilde{x}} \right) \cos( b\widetilde{y})
\end{equation}
generates the surface charge density $\sigma(\widetilde{y})$ via the relation 
(\ref{BCmodulated}),
\begin{equation} \label{sigmafinalLiou}
4\pi \ell_{\rm B}^2 \sigma(\widetilde{y}) =  \frac{2}{c} + \epsilon 
\left( b^2 + \frac{b}{c} + \frac{1}{c^2} \right) \cos( b\widetilde{y}) . 
\end{equation}
It is readily checked that the small $a$ limit of the non perturbative solution 
provided by Eq. \eqref{pot}, coincides with Eq. \eqref{phifinalLiou}.

Averaging equation \eqref{sigmafinalLiou} along the $\widetilde{y}$ axis over 
the period $2\pi/b$ implies that the parameter $c$ is related directly to 
the mean value of the surface charge density,
\begin{equation}
c = \frac{1}{2\pi\ell_{\rm B}^2\langle\sigma\rangle} . 
\end{equation}
To leading order in the smallness parameter $\epsilon$,
the contact relation takes the form
\begin{equation}
n(0,\tilde{y}) = \left[ 1 - \epsilon b^2 c \cos(b\tilde{y}) \right]
2\pi \ell_{\rm B} \sigma^2(\tilde{y}) ,
\end{equation}
with the prefactor smaller than 1 as was expected.

We recover here the same conclusion as above, although limited to 
a perturbative treatment: the corrugation (i.e. the $y$-dependence) of 
the surface charge $\sigma e$ is exponentially suppressed upon increasing 
the distance $x$ to the plate, see Eq. \eqref{phifinalLiou}
for the spatial dependence of the potential.
Besides, the connection between the period $2\pi/b$ of the pattern, 
and the pattern screening length $1/b$ is clearly apparent.

\renewcommand{\theequation}{4.\arabic{equation}}
\setcounter{equation}{0}

\section{Situations with added salt} \label{Sec4}
Now we turn to situations where a planar macroion is immersed in 
an infinite sea of electrolyte, playing the role of a salt reservoir 
and setting the Debye length $\kappa^{-1}$.

Let $x,y$ coordinates be measured in units of $1/\kappa$,
\begin{equation}
\widetilde{x} = \kappa x , \qquad \widetilde{y} = \kappa y .
\end{equation} 
We are looking for regular potential solutions of the 2D version of 
the PB equation (\ref{PB2})
\begin{equation} \label{shG}
\frac{\partial^2 \phi}{\partial\widetilde{x}^2} 
+ \frac{\partial^2 \phi}{\partial\widetilde{y}^2} = \sinh \phi ,
\end{equation}
the so-called 2D sinh-Gordon equation which is related to the better known 
2D sine-Gordon equation via the transformation $\phi\to i\phi$. 
The surface charge density, which in general depends on $\widetilde{y}$,
is again determined by the boundary condition
\begin{equation} \label{BCxy}
\frac{\partial\phi(\widetilde{x},\widetilde{y})}{
\partial\widetilde{x}}\Big\vert_{\widetilde{x}=0} = -
\frac{4\pi \ell_{\rm B} \sigma(\widetilde{y})}{\kappa} .
\end{equation} 
For completeness, we recall in Appendix \ref{app:salthomogeneous} the main 
results for a homogeneously charged plate. 

\subsection{2D Debye-H\"uckel solutions} \label{dh}
Unlike in the no-salt case, we start with a perturbative Debye-H\"uckel
(DH) treatment.
Within the DH approach, the linearization of $\sinh\phi \sim \phi$
in (\ref{shG}) leads to the Helmholtz equation
\begin{equation} \label{heq}
\frac{\partial^2 \phi_{\rm DH}}{\partial\widetilde{x}^2} + 
\frac{\partial^2 \phi_{\rm DH}}{\partial\widetilde{y}^2} = \phi_{\rm DH} . 
\end{equation}
Its solutions, which depend on both coordinates, can be obtained by using 
separation of variables:
\begin{equation}
\phi_{\rm DH} = \varphi(\widetilde{x}) \psi(\widetilde{y}) ,
\end{equation}
where $\varphi$ and $\psi$ obey the second-degree ordinary equations
\begin{equation}
\frac{1}{\varphi} \frac{d^2\varphi}{d \widetilde{x}^2} = \frac{1}{1-\gamma^2} ,
\qquad \frac{1}{\psi} \frac{d^2\psi}{d \widetilde{y}^2} 
= - \frac{\gamma^2}{1-\gamma^2} ,
\end{equation}
with the real parameter $\gamma\in (0,1)$.
In particular,
\begin{equation} \label{DHmod} 
\phi_{\rm DH} = c_1 \sin \left( \frac{\gamma\widetilde{y}}{\sqrt{1-\gamma^2}} 
\right) \exp\left( -\frac{\widetilde{x}}{\sqrt{1-\gamma^2}} \right) , 
\end{equation}
where $c_1$ is real.
This electrostatic potential is periodic along the $\widetilde{y}$-axis, 
with period (in units of the Debye length)
\begin{equation} \label{P}
{\cal P}(\gamma) = \frac{2\pi\sqrt{1-\gamma^2}}{\gamma} .
\end{equation}
Comparing the result (\ref{DHmod}) with the uniform DH solution (\ref{DHsol}), 
it is clear that any periodic modulation along the $\widetilde{y}$-axis 
implies a faster exponential decay in the $\widetilde{x}$ direction.
The decay rate along the $x$-axis depends on the period of the sine
function along the $y$-axis: larger period means smaller $\gamma$ and
consequently slower decay (the decay length is bounded from above by 
the Debye length, a value that is reached for an infinite period along $y$, 
i.e. with $\gamma=0$ \cite{rque102}). 
The form of the corresponding surface charge density follows from 
the boundary condition (\ref{BCxy}):
\begin{equation} \label{DHs}
\frac{4\pi \ell_{\rm B} \sigma_{\rm DH}(\widetilde{y})}{\kappa} 
= \frac{c_1}{\sqrt{1-\gamma^2}}
\sin \left( \frac{\gamma\widetilde{y}}{\sqrt{1-\gamma^2}} \right) .
\end{equation}

Since the Helmholtz equation (\ref{heq}) is linear, any superposition of 
particular solutions also is a solution:
\begin{eqnarray} 
\phi_{\rm DH} & = & c \exp\left(-\widetilde{x}\right) + \sum_n
c_n \sin \left( \frac{\gamma_n\widetilde{y}}{\sqrt{1-\gamma_n^2}} \right)
\nonumber \\ & & \times
\exp\left( -\frac{\widetilde{x}}{\sqrt{1-\gamma_n^2}} \right) , \label{DHmode} 
\end{eqnarray} 
where the parameters $\gamma_1<\gamma_2<\ldots<\gamma_N$ are from the
interval $(0,1)$, $c$ is any real constant and $c_1,\ldots,c_N$ are 
nonzero real constants.
The corresponding surface charge density $\sigma_{\rm DH}$ is given by
\begin{equation}
\frac{4\pi \ell_{\rm B} \sigma_{\rm DH}(\widetilde{y})}{\kappa} = c+\sum_n
\frac{c_n}{\sqrt{1-\gamma_n^2}}
\sin \left( \frac{\gamma_n\widetilde{y}}{\sqrt{1-\gamma_n^2}} \right) .
\end{equation}
Note that
\begin{equation}
\frac{4\pi \ell_{\rm B} \langle\sigma_{\rm DH}\rangle}{\kappa} = c .
\end{equation}
If $c\ne 0$, the  dominant term at large distances in (\ref{DHmode})
is the one with the uniform surface charge density equal to 
$\langle\sigma_{\rm DH}\rangle$.
If $c=0$, i.e. $\langle\sigma_{\rm DH}\rangle=0$, the dominant term corresponds 
to the smallest $\gamma_1$, i.e. to the largest period (\ref{P}). 

The extension of the DH formalism to general profiles of the surface charge 
varying along both $y$ and $z$ axis is straightforward.
On a general ground a Fourier mode of wave number $k$ for the charge pattern 
on the plate results in a far-field decay with a screening rate $\sqrt{1+k^2}$.
Hence, the smallest $k$ (the smallest $\gamma$) provides the mode that 
extends the furthest into the bulk.

Since the potential $\phi$ in the 2D Poisson equation (\ref{shG}) vanishes
at $\widetilde{x}\to\infty$, this equation can be linearized in the asymptotic
region (large $\widetilde x$) and its general solution is 
of type (\ref{DHmode}), which allows to define renormalized coefficients, 
following the uniform plate approach.
It is seen that the asymptotic potential is generically of the form 
$\exp(-\widetilde x)$, meaning that the surface corrugation is washed out 
with $\widetilde x$, and that the asymptotic decay is set by the Debye length.
At finite distance from the wall, surface charge modulations with various 
periods influence each other due to the nonlinearity of the sinh-Gordon 
equation. 
One may surmise here that the above generic scenario holds provided 
$\langle\sigma\rangle\ne 0$. 
The situation with $\langle\sigma\rangle=0$ is more subtle to analyze;
an explicit case is worked out below.
Finally, we emphasize that the phenomenon of saturation can be documented
on exactly solvable cases; it corresponds to the fact that a divergent
surface charge may nevertheless yield a finite potential at all points 
outside the charged body creating the field \cite{TT03}.     

\subsection{Soliton solutions of 2D Poisson-Boltzmann/sinh-Gordon equation}
All solutions of the 2D equation (\ref{shG}) are available due 
to the existence of B\"acklund transformation which reduces the second-order 
differential equation (\ref{shG}) to a couple of the first-order ones 
\cite{Rogers82}.
The simplest soliton one-particle solutions, formulated standardly within 
the related 2D sine-Gordon theory, is used to generate via 
the B\"acklund transformation solutions with higher number of soliton 
``particles'' \cite{Hirota92,Samaj13}.

The one-soliton solution has the form
\begin{widetext}
\begin{equation} \label{onesoliton}
\phi = 2 \ln \left\{ \frac{\exp{\left[ (\widetilde{x}+a)/\sqrt{1+\gamma^2}
\right]} + \xi \exp{\left[-\gamma (\widetilde{y}+b)/\sqrt{1+\gamma^2}\right]}}{
\exp{\left[(\widetilde{x}+a)/\sqrt{1+\gamma^2}\right]}
- \xi \exp{\left[-\gamma (\widetilde{y}+b)/\sqrt{1+\gamma^2}\right]}} \right\} ,
\end{equation}
\end{widetext}
where the coordinate shifts $a$ and $b$ are arbitrary (they only renormalize
$\xi>0$) and the parameter $\gamma$ is real.
There always exist negative values of $\widetilde{y}$ such that the denominator 
of the fraction under logarithm is equal to 0 or negative, which is physically 
unacceptable. 
To put it differently, Eq. \eqref{onesoliton} leads to physically reasonable 
solution in the upper quadrant only ($x\geq 0; y\geq 0$) and we do not dwell 
further on its properties.
The only exception is when $\gamma=0$ for which (\ref{onesoliton}) 
(with $a=0$) reduces to the uniformly charged plate solution (\ref{homosol}). 

We turn to the two-soliton solutions of the sinh-Gordon equation, that can be 
written as a formal generalization of the one-soliton result 
(\ref{onesoliton}):
\begin{equation} \label{ansatze}
\phi(\tilde{x},\tilde{y}) = 2 \ln 
\left[ \frac{f(\widetilde{x})+g(\widetilde{y})}{
f(\widetilde{x})-g(\widetilde{y})} \right] .
\end{equation}
The function $g(\tilde{y})$ should obey the differential equation
\begin{equation} \label{gy1}
[g'(\widetilde{y})]^2 = A g^4(\widetilde{y}) - B g^2(\widetilde{y}) + C 
\end{equation}
with some as-yet undetermined real coefficients $A$, $B$ and $C$.
The derivation of this equation with respect to $\widetilde{y}$ yields
\begin{equation} \label{gy2}
g''(\widetilde{y}) = 2 A g^3(\widetilde{y}) - B g(\widetilde{y}) .
\end{equation}
Similarly, the function $f(\widetilde{x})$ satisfies the equation
\begin{equation} \label{fx1}
[f'(\widetilde{x})]^2 = A' f^4(\widetilde{x}) - B' f^2(\widetilde{x}) + C' 
\end{equation}
with some other real coefficients $A'$, $B'$ and $C'$.
As before, differentiating this equation with respect to $\widetilde{x}$ yields
\begin{equation} \label{fx2}
f''(\widetilde{x}) = 2 A' f^3(\widetilde{x}) - B' f(\widetilde{x}) .
\end{equation}
Inserting the ansatz (\ref{ansatze}) into the sinh-Gordon equation (\ref{shG})
and using the relations (\ref{gy1})--(\ref{fx2}) it can be shown that 
the functions $f(x)$ and $g(y)$ provide the solution of (\ref{shG}) if
\begin{equation}
A' = - A , \qquad B' = - (B+1) , \qquad C' = -C .
\end{equation}
For a special choice of the coefficients
\begin{eqnarray}
& & A = A' = 0 , \qquad B = C = \frac{\gamma^2}{1-\gamma^2} , \nonumber \\
& & B' =  - \frac{1}{1-\gamma^2} , \qquad C' = -\frac{\gamma^2}{1-\gamma^2} 
\end{eqnarray}
with the real parameter 
\begin{equation}
0<\gamma<1 ,
\end{equation} 
the $f$ and $g$ functions are obtained as follows
\begin{equation}
f = \gamma \cosh\left( \frac{\widetilde{x}}{\sqrt{1-\gamma^2}}+a 
\right) , \quad
g = \sin \left( \frac{\gamma\widetilde{y}}{\sqrt{1-\gamma^2}} \right) , 
\end{equation}
where $a$ is a real positive number.
The resulting potential is
\begin{equation} \label{respot}
\phi = 2 \ln \left[ \frac{\gamma \cosh\left( 
\frac{\widetilde{x}}{\sqrt{1-\gamma^2}}+a \right) + \sin \left( 
\frac{\gamma\widetilde{y}}{\sqrt{1-\gamma^2}} \right)}{\gamma \cosh\left( 
\frac{\widetilde{x}}{\sqrt{1-\gamma^2}} + a \right) - \sin \left( 
\frac{\gamma\widetilde{y}}{\sqrt{1-\gamma^2}} \right)} \right] .
\end{equation}
Keeping in mind that $x\ge 0$, the inequality 
\begin{equation} \label{restr} 
\gamma \cosh a > 1 
\end{equation}
must hold in order to avoid the singularity in $\phi$.
This inequality is equivalent to 
\begin{equation} \label{condition}
a > a_c = \ln\left[\frac{1}{\gamma}\left( 1+\sqrt{1-\gamma^2} 
\right)\right] .
\end{equation}

\begin{figure}
\begin{center}
\includegraphics[width=0.45\textwidth,clip]{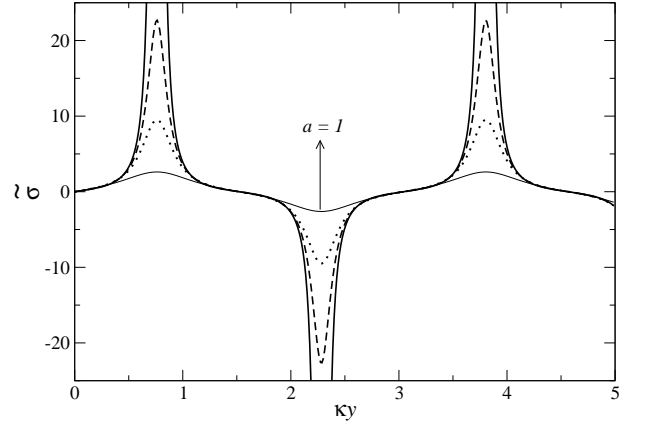}
\caption{Plot of the dimensionless surface charge 
$\widetilde \sigma = \pi l_B \sigma/\kappa$ stemming from 
Eq. \eqref{sigmaperiod} for $\gamma=0.9$, as a function of 
$\widetilde y = \kappa y$.
For this situation, the period of the charge pattern is ${\cal P} \simeq 3.04$. 
The thick continuous, dashed, dotted and thin continuous curves are for 
$a=a_c$, $a=0.52$, $0.6$ and 1, respectively.
The critical $a$ associated to the constraint \eqref{condition} is 
$a_c \simeq 0.467$, it leads to a locally diverging surface charge.}
\label{fig:surfacecharge}
\end{center}
\end{figure}

\begin{figure}
\begin{center}
\includegraphics[width=0.45\textwidth,clip]{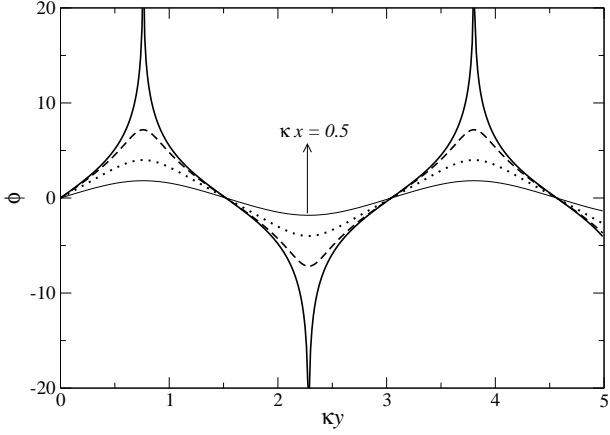}
\caption{Electrostatic potential profiles as given by \eqref{respot}. 
The plot shows the $y$ dependence (parallel to the charged plate), 
for different distances to the plates: $\gamma=0.9$ as in 
Fig. \ref{fig:surfacecharge} with $a=a_c$. 
The thick continuous, dashed, dotted and thin continuous curves are for 
$\widetilde x = \kappa x = 0$, $0.05$, $0.2$ and $0.5$, respectively.}
\label{fig:pot}
\end{center}
\end{figure}

Using the boundary condition (\ref{BCxy}), the surface charge density 
is given by 
\begin{equation} \label{sigmaperiod}
\sigma = \frac{\kappa}{\pi\ell_{\rm B}} \frac{\gamma \sinh a}{\sqrt{1-\gamma^2}} 
\frac{\sin\left( \frac{\gamma\widetilde{y}}{\sqrt{1-\gamma^2}}\right)}{\gamma^2 
\cosh^2 a -\sin^2\left( \frac{\gamma\widetilde{y}}{\sqrt{1-\gamma^2}} \right)}. 
\end{equation}
This function is periodic with the period ${\cal P}(\gamma)$ given by (\ref{P}).
The parameter $a$, constrained by the inequality (\ref{restr}), controls
the amplitude of oscillations which is enhanced when $\gamma \cosh a$ is 
close to 1.  
Since $\sigma(\widetilde{y}) = -\sigma(-\widetilde{y})$, the mean value of the 
surface charge density over the period vanishes,
\begin{equation}
\langle \sigma \rangle = 0 .
\end{equation}
The behavior of the surface charge is shown in Fig. \ref{fig:surfacecharge}.
For $a=a_c$, the surface charge is divergent at specific points. 
Yet, the electrostatic potential is regular for $x>0$, see Fig. \ref{fig:pot}.
For $x=0$, the potential exhibits a diverging tip at the points 
where $\sigma$ diverges.

The reduced potential decays at large distances from the wall as
\begin{equation} \label{phiasymp}
\phi(\widetilde{x},\widetilde{y}) \mathop{\sim}_{\widetilde{x}\to\infty} 
\frac{8 e^{-a}}{\gamma} 
\sin\left( \frac{\gamma\widetilde{y}}{\sqrt{1-\gamma^2}}\right) 
\exp\left( - \frac{\widetilde{x}}{\sqrt{1-\gamma^2}} \right) .
\end{equation}  
It is seen that the surface charge density (\ref{sigmaperiod}), which 
is periodic function of $\widetilde{y}$ with a relatively complicated Fourier
series, implies at asymptotic distances from the wall the potential of 
the DH form (\ref{DHmod}) as was expected.
The exact relationship to the DH theory can be documented by considering
the limit $a\to\infty$ of the surface charge density (\ref{sigmaperiod}):
\begin{equation}
\frac{4\pi \ell_{\rm B} \sigma(\widetilde{y})}{\kappa} 
\sim \frac{8}{\gamma} \frac{1}{\sqrt{1-\gamma^2}} e^{-a}
\sin \left( \frac{\gamma\widetilde{y}}{\sqrt{1-\gamma^2}} \right) ,
\end{equation}
which corresponds to the DH surface charge density (\ref{DHs}) with
$c_1 = 8 e^{-a}/\gamma \ll 1$.
The DH potential (\ref{DHmod}) is then equivalent to our asymptotic
potential (\ref{phiasymp}).
Eq. \eqref{phiasymp} indicates that the asymptotic screening length
is (in units of the Debye length) $\ell=\sqrt{1-\gamma^2}$. 
For the $\gamma$-parameter of Fig.  \ref{fig:pot}, this yields a length 
$\ell\simeq 0.43$.
This is compatible with the data shown in Fig. \ref{fig:pot}, where it is seen 
that for $\widetilde x= \kappa x = 0.5$ already, $\phi$ exhibits significantly 
reduced oscillations. 
The linear response regime, where $\phi$ is everywhere smaller than 1, 
is reached for $\widetilde x >0.76 $.

The particle species densities, given by $n_{\pm} = (n/2) e^{\mp\phi}$, read as
\begin{eqnarray}
n_+ & = & \frac{n}{2} \left[ \frac{\gamma \cosh\left( 
\frac{\widetilde{x}}{\sqrt{1-\gamma^2}} +a \right) - \sin \left( 
\frac{\gamma\widetilde{y}}{\sqrt{1-\gamma^2}} \right)}{\gamma \cosh\left( 
\frac{\widetilde{x}}{\sqrt{1-\gamma^2}}+a \right) + \sin \left( 
\frac{\gamma\widetilde{y}}{\sqrt{1-\gamma^2}} \right)} \right]^2 , \nonumber \\
n_- & = & \frac{n}{2} \left[ \frac{\gamma \cosh\left( 
\frac{\widetilde{x}}{\sqrt{1-\gamma^2}}+a \right) + \sin \left( 
\frac{\gamma\widetilde{y}}{\sqrt{1-\gamma^2}} \right)}{\gamma \cosh\left( 
\frac{\widetilde{x}}{\sqrt{1-\gamma^2}} + a\right) - \sin \left( 
\frac{\gamma\widetilde{y}}{\sqrt{1-\gamma^2}} \right)} \right]^2 . \nonumber \\
& &
\end{eqnarray}
At each distance from the wall $\widetilde{x}$, the particle species densities 
fulfill the equality 
$n_+(\widetilde{x},\widetilde{y}) = n_-(\widetilde{x},-\widetilde{y})$, 
so the integral over the charge density vanishes, i.e. 
\begin{equation} \label{rho}
\langle \rho(\widetilde{x}) \rangle = 0 \quad 
\mbox{for each $\widetilde{x}\in [0,\infty)$.}
\end{equation}
For large distances from the wall, the particle charge density decays as
\begin{equation} \label{rhoasymp}
\rho(\widetilde{x},\widetilde{y}) \mathop{\sim}_{\widetilde{x}\to\infty}
\frac{8e n e^{-a}}{\gamma}
\sin \left( \frac{\gamma\widetilde{y}}{\sqrt{1-\gamma^2}}\right) 
\exp\left( - \frac{\widetilde{x}}{\sqrt{1-\gamma^2}} \right) .
\end{equation}  

The total particle density at the wall
$n(0,\widetilde{y}) \equiv n_+(0,\widetilde{y}) + n_-(0,\widetilde{y})$ 
satisfies a local relation of type (\ref{localrel}) 
\begin{equation} \label{nsalt}
n(0,\widetilde{y}) - n \, = \,  \frac{1-\gamma^2}{\tanh^2 a} \,\,
2\pi \ell_{\rm B} \sigma^2(\widetilde{y}) .
\end{equation}
Using the restriction on the $a$ parameter (\ref{condition}) it can
be shown that the prefactor
\begin{equation}
\frac{1-\gamma^2}{\tanh^2 a} < 1 ,
\end{equation}
in agreement with the general theory presented in Sec. \ref{Sec2}.
It is easy to show that the contact relations (\ref{P2}) and (\ref{C3}) 
are satisfied.
The mean value (over the period) of the total particle density as 
the function of the distance from the wall $\widetilde{x}$ behaves as
\begin{equation}
\frac{\langle n(\widetilde{x})\rangle}{n} - 1 = \frac{4 
\left[\gamma \cosh\left( \frac{\widetilde{x}}{\sqrt{1-\gamma^2}}+
a\right)\right]}{\left\{ \left[\gamma \cosh\left( 
\frac{\widetilde{x}}{\sqrt{1-\gamma^2}}+a\right) \right]^2 - 1 \right\}^{3/2}} .
\end{equation}
At asymptotically large distances $\widetilde{x}$, one has
\begin{equation}
\frac{\langle n(\widetilde{x})\rangle}{n} - 1 \mathop{\sim}_{\widetilde{x}\to\infty}
\frac{16 e^{-2a}}{\gamma^2} 
\exp\left( -\frac{2\widetilde{x}}{\sqrt{1-\gamma^2}} \right) ,
\end{equation}
or a more detailed asymptotic relation
\begin{eqnarray}
\frac{n(\widetilde{x},\widetilde{y})}{n} - 1 
& \displaystyle{\mathop{\sim}_{\widetilde{x}\to\infty}} & \frac{32 e^{-2a}}{\gamma^2} 
\sin^2\left( \frac{\gamma\widetilde{y}}{\sqrt{1-\gamma^2}}\right) \nonumber \\
& & \times \exp\left( -\frac{2\widetilde{x}}{\sqrt{1-\gamma^2}} \right) . 
\end{eqnarray}
Comparing this formula with its analogue for the particle charge density
(\ref{rhoasymp}), we see that the approach of the particle number to 
its bulk value is faster by a factor $2$ in the exponential.

\begin{figure}
\begin{center}
\includegraphics[width=0.45\textwidth,clip]{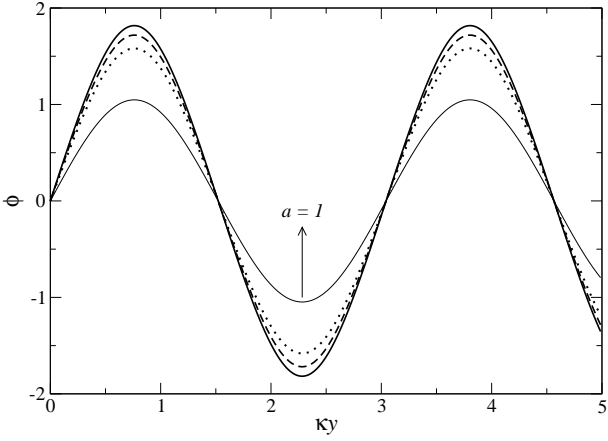}
\caption{Electrostatic potential at a fixed distance from the wall, 
$\widetilde x=0.5$, for the various charge patterns shown in 
Fig. \ref{fig:surfacecharge}. 
The thick continuous, dashed, dotted and thin continuous curves are for 
$a=a_c$, $a=0.52$, $0.6$ and 1 respectively, with $a_c \simeq 0.467$.}
\label{fig:pot_sat}
\end{center}
\end{figure}

Finally, let the amplitude of the surface charge density (\ref{sigmaperiod})
go to infinity, i.e. $\gamma\cosh a = 1$, which defines $a_c$. 
Equivalently,
\begin{equation}
e^{a_c} = \frac{1}{\gamma} \left( 1 + \sqrt{1-\gamma^2} \right) .
\end{equation}
Considering this relation in the asymptotic decay of the potential
(\ref{phiasymp}), the prefactor 
\begin{equation} \label{prefactor}
\frac{8 \, e^{-a}}{\gamma} = \frac{8}{1 + \sqrt{1-\gamma^2}}
\end{equation}
becomes finite which is evidence for the saturation phenomenon \cite{TT03}.
Note that the saturated prefactor depends on $\gamma$ and its value
ranges between $4$ for $\gamma\to 1$ and $8$ for $\gamma\to 0$. 
This leads to the remark that the situation leading to the 
most enhanced large-distance potential is when $\gamma \to 0$, meaning that the
period of the charge pattern diverges. 
We have already met earlier this feature. 
Figure \ref{fig:pot_sat} illustrates saturation of the electrostatic signature,
for the different charge patterns presented in Fig. \ref{fig:surfacecharge}.
It is observed that for $a\leq 0.6$, the potential at the chosen distance 
from the plate ($\widetilde x = 0.5$), depends quite weakly on $a$, 
while the surface charge evolves from strongly modulated at $a=0.6$, 
to locally divergent at $a=a_c$. 
Besides, the divergent surface charge for $a=a_c$ yields a well behaved
potential.
The potential for $a=1$ is distinct from the other three, since it corresponds 
to too weak modulation.

A natural next step is to proceed to many-soliton solutions, using e.g.
a simplified Hirota's method \cite{Wazwaz12}.
The problem is that the transition from the sine-Gordon to sinh-Gordon theories
via the transformation $\phi\to i\phi$ converts  regular solutions to 
unacceptable singular ones. 

\subsection{A perturbative solution of the Poisson-Boltzmann/sinh-Gordon
equation} \label{pp}
The above non-perturbative solution of the 2D Poisson-Boltzmann equation for 
the potential (\ref{respot}) contains two independent parameters 
$\gamma\in (0,1)$ and $a$, constrained by (\ref{restr}).
By varying these parameters one can obtain a number of various forms 
of the corresponding surface charge density (\ref{sigmaperiod}), however all
forms have the common property that $\langle\sigma\rangle = 0$.
In analogy with the 2D Liouville equation in Sec. \ref{ppp}, 
we propose in what follows a perturbative treatment of the 2D problem
around the full solution of the uniform surface charge density,
by considering a small periodic modulation.
We recall that in the standard DH approach the {\em whole} potential is taken 
as a small quantity in which case the sinh function can be linearized.
The present treatment thus differs from the linear response derived in section 
\ref{dh}.

Let us add to the uniform potential solution (\ref{homosol})
an infinitesimal perturbation $\epsilon f(\widetilde{x},\widetilde{y})$ 
with $\epsilon\ll 1$:
\begin{equation}
\phi(\widetilde{x},\widetilde{y}) = \phi_0(\widetilde{x}) 
+ \epsilon f(\widetilde{x},\widetilde{y}) , \qquad \phi_0 =  2 
\ln\left( \frac{e^{\widetilde{x}}+\xi}{e^{\widetilde{x}}-\xi} \right) .
\end{equation}
The parameter $\xi$ is as-yet unspecified, and will be related to 
the surface charge density.
Inserting this ansatz into the 2D sinh-Gordon equation (\ref{shG}), 
the function $f(x,y)$ must obey
\begin{equation} \label{feq}
\frac{\partial^2 f}{\partial\widetilde{x}^2} + 
\frac{\partial^2 f}{\partial\widetilde{y}^2} = f \cosh\phi_0 ,
\end{equation}
where
\begin{eqnarray}
\cosh\phi_0(\widetilde{x}) & = & \frac{1}{2} \left[
\left( \frac{e^{\widetilde{x}}+\xi}{e^{\widetilde{x}}-\xi} \right)^2
+ \left( \frac{e^{\widetilde{x}}-\xi}{e^{\widetilde{x}}+\xi} \right)^2
\right] \nonumber \\ & = &
1 + \frac{8\xi^2 e^{-2\widetilde{x}}}{\left( 1-\xi^2 e^{-2\widetilde{x}}\right)^2} .
\end{eqnarray}
Note that Eq. (\ref{feq}) is in fact the linearized DH version of the
sinh-Gordon equation with the position-dependent 
$\kappa(\widetilde{x})= \sqrt{4\pi\ell_{\rm B}n(\widetilde{x})}$, where
$n(\widetilde{x})$ is the standard PB total particle density for the uniformly
charged plate.

Using separation of variables 
\begin{equation}
f(\widetilde{x},\widetilde{y}) = \varphi(\widetilde{x}) \psi(\widetilde{y})
\end{equation}
in Eq. (\ref{feq}), the $\varphi$ and $\psi$ functions must obey 
\begin{equation} \label{twode}
\frac{1}{\varphi} \frac{d^2\varphi}{d \widetilde{x}^2} = 
\frac{1}{1-\gamma^2} + \frac{8\xi^2 e^{-2\widetilde{x}}}{\left( 
1-\xi^2 e^{-2\widetilde{x}}\right)^2} , \qquad
\frac{1}{\psi} \frac{d^2\psi}{d \widetilde{y}^2} 
= - \frac{\gamma^2}{1-\gamma^2} .
\end{equation}
The solution for $\psi$ reads as
\begin{equation}
\psi = \sin \left( \frac{\gamma\widetilde{y}}{\sqrt{1-\gamma^2}} \right) .
\end{equation}
The solution for $\varphi$ is searched as a series
\begin{equation}
\varphi = e^{-\widetilde{x}/\sqrt{1-\gamma^2}}
\sum_{n=0}^{\infty} c_n \xi^{2n} e^{-2n\widetilde{x}} . 
\end{equation}
Inserting this series into Eq. (\ref{twode}) implies a recurrent
scheme for the coefficient
\begin{equation} \label{rec}
\frac{1}{2} n \left( n + \frac{1}{\sqrt{1-\gamma^2}} \right) c_n
= \sum_{j=0}^{n-1} (n-j) c_j ,
\end{equation}
where $n=1,2,\ldots$ and $c_0$ is free.
It is straightforward to verify that the constant series
\begin{equation}
c_n = \frac{2\sqrt{1-\gamma^2}}{1+\sqrt{1-\gamma^2}} c_0
\end{equation}
solves the recursion (\ref{rec}).
Setting $c_0=1$, $\varphi$ is found to be
\begin{equation}
\varphi = e^{-\widetilde{x}/\sqrt{1-\gamma^2}}
\left( 1 + \frac{2\sqrt{1-\gamma^2}}{1+\sqrt{1-\gamma^2}}
\frac{\xi^2 e^{-2\widetilde{x}}}{1-\xi^2 e^{-2\widetilde{x}}} \right) .
\end{equation}

The total potential reads as
\begin{eqnarray}
\phi & = & 2 \ln\left( \frac{e^{\widetilde{x}}+\xi}{e^{\widetilde{x}}-\xi} 
\right) + \epsilon 
\sin \left( \frac{\gamma\widetilde{y}}{\sqrt{1-\gamma^2}} \right) 
e^{-\widetilde{x}/\sqrt{1-\gamma^2}}   \nonumber \\ & & \times 
\left( 1 + \frac{2\sqrt{1-\gamma^2}}{1+\sqrt{1-\gamma^2}}
\frac{\xi^2 e^{-2\widetilde{x}}}{1-\xi^2 e^{-2\widetilde{x}}} \right) . 
\label{phifinal}
\end{eqnarray}
The corresponding surface charge density $\sigma(\widetilde{y})$,
generated via the relation (\ref{BCxy}), takes the form
\begin{eqnarray}
\frac{4\pi \ell_{\rm B} \sigma(\widetilde{y})}{\kappa} & = & 
\frac{4\xi}{1-\xi^2} + \epsilon \sin 
\left( \frac{\gamma\widetilde{y}}{\sqrt{1-\gamma^2}} \right) 
\nonumber \\ & & \times
\frac{(1+\xi^2)^2+\sqrt{1-\gamma^2}(1-\xi^4)-2\gamma^2\xi^2}{\sqrt{1-\gamma^2}
(1+\sqrt{1-\gamma^2})(1-\xi^2)^2} . \nonumber \\ & & \label{sigmafinal}
\end{eqnarray}
Averaging this equation over the period leads to
\begin{equation}
\frac{4\pi \ell_{\rm B} \langle\sigma\rangle}{\kappa} \equiv a
= \frac{4\xi}{1-\xi^2} ,
\end{equation}
where $\xi$ as the function of $a$ is expressed in Eq. (\ref{xi}).

It stands to reason that in the limit $\langle \sigma\rangle\to 0$
$(\xi\to 0)$ our equations (\ref{phifinal}) and (\ref{sigmafinal})
reduce to their DH counterparts (\ref{DHmod}) and (\ref{DHs}).
The conclusions of previous sections on screening lengths and periodicity
are unaltered.

\renewcommand{\theequation}{5.\arabic{equation}}
\setcounter{equation}{0}

\section{Conclusion} \label{Sec5}
This article was devoted to deriving new analytical solutions to the 
Poisson-Boltzmann theory, that describes equilibrium electric double-layers 
around charged macromolecules. 
We have addressed deionized situations (counterions only, also known as 
salt-free) and others when the double-layer is in equilibrium with 
a bulk of salt, playing the role of a reservoir.
Previously known solutions pertain to uniformly charged macromolecules,
and we focused on models with inhomogeneous surface charge densities, 
referred to as patterns.
In doing so, generic effects of screening emerge.

The no-salt case was solved in Sec. \ref{Sec3}, taking advantage of known 
results for the 2D Liouville equation (\ref{PBmodulated}) for the mean (reduced)
potential $\phi$.
All solutions of this equation are known, see relations 
(\ref{cv})-(\ref{constraint}).
Once a solution for $\phi$ is chosen, the corresponding surface charge density 
is generated in an inverse way from the boundary condition (\ref{BCmodulated}).
The problem with these solutions is that the great majority of them have
singularities (divergencies) in the particle region $\Lambda$ and/or
the nonvanishing derivative of $\phi$ with respect to $x$ at $x\to\infty$
which corresponds to unphysical non-neutral charge systems.
A generic feature in the no-salt case is that a periodic charge pattern 
(with non-vanishing mean) is screened exponentially in the planar case. 
This may be surprising since the counterion density, a measure of charge 
screening, decays algebraically, as the inverse squared distance to the plate.
One should thus distinguish charge screening (the recovery of a neutral 
system at large distance), from heterogeneity screening (the loss of 
charge/potential patterning). 
The situation for a charged cylinder differs, in the sense that pattern 
screening becomes algebraic. 
This can be rationalized by the planar to cylindrical mapping presented 
in section \ref{l}, which highlights the Cartesian/cylindrical coordinates
correspondence $\widetilde{x} \leftrightarrow \ln \widetilde{r}$,
$\widetilde{y} \leftrightarrow \varphi$. 
More precisely, a charge pattern on the cylinder with angular period 
$2 \pi/b$ where $b$ is some integer, has a signature in the potential that 
decays as the inverse power-law $\widetilde{r}^{-b}$. 
This is superimposed to the global decay of potential/density away from
the charged cylinder, that reduces at large distance to that of a cylinder 
without any charge pattern.

For system with added salt, our main results pertain to planar interfaces 
with periodic charge patterns. 
The planar to cylindrical correspondence is lost. 
This discussion is developed in Sec. \ref{Sec4}.
The linearized DH approach, based on the Helmholtz equation (\ref{heq}),
provides the modulated solutions of type (\ref{DHmod}) where the decay
rate along the $x$-axis is directly related to the period of the sine
function along the $y$-axis.
The nonlinear Poisson-Boltzmann approach is associated with 
the 2D sinh-Gordon equation (\ref{shG}).
This equation is integrable and possesses a number of many-soliton 
solutions, but practically all solutions suffer from singularities within 
the space $\Lambda$ occupied by the particles.
An exception is represented by the 2-soliton solution for the potential 
(\ref{respot}) which implies the periodic surface charge density 
(\ref{sigmaperiod}) with zero mean.
It is interesting that this relatively complicated solution provides
the local contact relation (\ref{nsalt}).
The phenomenon of saturation \cite{TT03,Trizac02} is documented on this model:
increasing the amplitude of periodic oscillations of the surface charge density
to infinity implies the asymptotic decay of the potential (\ref{phiasymp})
with the finite prefactor (\ref{prefactor}).
To understand also the systems with a nonzero mean of the surface charge 
density, we constructed in Sec. \ref{pp} a perturbative treatment of the 2D
sinh-Gordon equation in infinitesimal periodic modulations of the uniform
surface charge density, in close analogy with the 2D Liouville equation.
Moreover, we have found that the signature of surface charge pattern extends 
all the more into the bulk electrolyte as the associated period of 
the pattern is large. 
The connection between the period $\cal P$ of the charge pattern and 
the screening length $\ell$ reads
\begin{equation}
 \frac{\ell}{\cal P} \,=\, \frac{1}{\sqrt{ \kappa^2{\cal P}^2 + (2\pi)^2 } }.
\end{equation}
For small period $\kappa {\cal P} \ll 1$, we have $\ell \sim {\cal P}/(2\pi)$.
Increasing $\cal P$, the screening length increases as well, and saturates to 
$\kappa^{-1}$ for large periods.

A general analysis of the statistical quantities at the wall contact was 
the subject of Sec. \ref{Sec2}.
Using the pressure tensor, we have derived the integral
constraint for the local pressure given by Eqs. (\ref{P1}) and (\ref{C1}) 
which relates the surface charge density and the statistical quantities at 
the wall contact, namely the particle density and the parallel component of 
the electric field.
This integral constraint was verified to be true for every exactly
solvable model.
The inequality (\ref{ineq1}) consequently applies.
An important feature for the no-salt case is the confirmation of 
the enhancement of the counterion density at the wall in comparison
with the uniform case.
It is possible that the established upper bound of the mean contact 
particle density in Eq. (\ref{lowerupper}) is of general validity. 

Variations of the surface charge density studied in this paper were restricted
to one direction, so that the proposed solutions depend on two coordinates, 
and not three.
It would be useful to have exactly solved models for more general profiles
of the surface charge varying along both $y$ and $z$ axis, but this
requires the solution of 3D versions of the Liouville or sinh-Gordon equations.
Although a 3D B\"acklund transformation has already been proposed for
the Liouville equation \cite{Leibbrandt80,Huang06}, the exact solutions
seem out of reach. 

\emma{
Finally, while we focussed on one macroion features, it would be relevant 
to study macroion-macroion interactions within this formalism, and to 
compare to known results.   It was indeed shown recently that nano-patterned 
surfaces exhibit an interaction force that strongly depends on the 
alignement between charged domains, and of the domain size
\cite{Bakhshandeh18,comment2001}. Work along these lines is in progress.
}

\begin{acknowledgments}
L. \v{S}. is grateful to LPTMS for hospitality. 
The support received from the project EXSES APVV-16-0186 and VEGA Grant 
No. 2/0003/18 is acknowledged. 
The work was funded by the European Union's Horizon 2020 research 
and innovation programme under ETN grant 674979-NANOTRANS.
\end{acknowledgments}

\appendix

\renewcommand{\theequation}{A.\arabic{equation}}
\setcounter{equation}{0}

\section{Rederivation of the integral pressure relation}
\label{app:Rederivation}
We show here how to recover Eqs. (\ref{P1}) and (\ref{C1}) by direct use of 
the PB equation (\ref{PB1}). 
We consider the counterion only situation.
Multiplying the PB equation by $\partial\phi/\partial x$
and integrating over $x$ from $0$ to $\infty$, we obtain
\begin{eqnarray}
4\pi\ell_{\rm B} n(0,y,z)-\frac{1}{2} 
\left( \frac{\partial\phi}{\partial x} \right)^2\Big\vert_{x=0}
\phantom{aaaaa} \nonumber \\
+ \int_0^{\infty} dx \frac{\partial\phi}{\partial x} 
\frac{\partial^2\phi}{\partial y^2}
+ \int_0^{\infty} dx \frac{\partial\phi}{\partial x} 
\frac{\partial^2\phi}{\partial z^2} = 0 .
\end{eqnarray}
Next we use the boundary condition (\ref{BCreduced}) at $x=0$, divide the 
above equation by $4\pi\ell_{\rm B}$ and finally integrate it 
over $y$ and $z$ from $-\infty$ to $\infty$, to get
\begin{eqnarray} \label{PP}
\int_{-\infty}^{\infty} dy \int_{-\infty}^{\infty} dz 
\left[ n(0,y,z) - 2\pi\ell_{\rm B} \sigma^2(y,z) \right]
\phantom{aaaaa} \nonumber \\ + \frac{1}{4\pi\ell_{\rm B}}  
\int_0^{\infty} dx \int_{-\infty}^{\infty} dy \int_{-\infty}^{\infty} dz 
\frac{\partial\phi}{\partial x} \frac{\partial^2\phi}{\partial y^2}
\nonumber \\ + \frac{1}{4\pi\ell_{\rm B}}  
\int_0^{\infty} dx \int_{-\infty}^{\infty} dy \int_{-\infty}^{\infty} dz 
\frac{\partial\phi}{\partial x} \frac{\partial^2\phi}{\partial z^2} = 0 .
\nonumber \\
\end{eqnarray}
Using integrations by parts with the neglection of boundary terms at infinity
in the $(x,y)$-subspace, we get the following equivalence of integrals:
\begin{eqnarray}
\int_{-\infty}^{\infty} dy \left( \frac{\partial\phi}{\partial y} 
\right)^2\Big\vert_{x=0} & = & - \int_{-\infty}^{\infty} dy \int_0^{\infty} dx
\frac{\partial}{\partial x} \left( \frac{\partial\phi}{\partial y} \right)^2 
\nonumber \\ & = & - 2 \int_0^{\infty} dx \int_{-\infty}^{\infty} dy 
\frac{\partial^2\phi}{\partial x\partial y} \frac{\partial\phi}{\partial y} 
\nonumber \\ & = & 2 \int_0^{\infty} dx \int_{-\infty}^{\infty} dy 
\frac{\partial\phi}{\partial x} \frac{\partial^2\phi}{\partial y^2} . 
\nonumber \\ & &
\end{eqnarray}
Proceeding similarly in the $(x,z)$-subspace results in
\begin{equation}
\int_{-\infty}^{\infty} dz \left( \frac{\partial\phi}{\partial z} 
\right)^2\Big\vert_{x=0} = 2 \int_0^{\infty} dx \int_{-\infty}^{\infty} dz 
\frac{\partial\phi}{\partial x} \frac{\partial^2\phi}{\partial z^2} . 
\end{equation}
Inserting the last two integral equalities into (\ref{PP}), we arrive
at the contact relation given by Eqs. (\ref{P1}) and (\ref{C1}).

\renewcommand{\theequation}{B.\arabic{equation}}
\setcounter{equation}{0}

\section{Derivation of non-neutral solutions} \label{app:nonneutral}
Choosing in (\ref{phisimplest}) the parameters
\begin{equation}
c_1 = \frac{1}{2c} , \quad c_2 = - \frac{c}{2} + \frac{a^2}{2c} , \quad
\alpha = \frac{a}{2c} , \quad \beta = 0 , 
\end{equation}
which fulfill the constraint (\ref{constraintsimplest}), we obtain
\begin{equation}
\phi = - 2 \ln \left\{ \frac{1}{2c} 
\left[ (\widetilde{x}+a)^2 + \widetilde{y}^2 \right] - \frac{c}{2} \right\} . 
\label{eq:phisol}
\end{equation}
To ensure that the expression under logarithm is positive at any point 
in $\Lambda$, it is necessary that
\begin{equation}
0 < c < a .
\end{equation}
The boundary condition (\ref{BCmodulated}) yields the surface charge density
on the plate at $x=0$:
\begin{equation} \label{sigmaloc}
\sigma = \frac{1}{\pi\ell_{B}^2} \frac{a}{a^2-c^2+\widetilde{y}^2} .
\end{equation}
It is maximal at $\widetilde{y}=0$ and monotonously decays to zero for 
$\widetilde{y}\to\pm\infty$.
The corresponding density profile follows from Eq. (\ref{nmodulated}),
\begin{equation}
n = \frac{2 c^2}{\pi \ell_{B}^3} 
\frac{1}{\left[ (\widetilde{x}+a)^2 - c^2 + \widetilde{y}^2 \right]^2} .
\label{eq:prof}
\end{equation} 
There exists a non-trivial local relation between the particle density 
at the wall and the surface charge density
\begin{equation} \label{localrel1}
n(0,\widetilde{y}) = \left( \frac{c}{a} \right)^2 2\pi \ell_{\rm B} 
\sigma^2(\widetilde{y}) 
\end{equation}
which is of type (\ref{localrel}) with the prefactor $(c/a)^2<1$,
in agreement with the theory developed in Sec. \ref{Sec2}. 
The integral contact relation, given by Eqs. (\ref{P2}) and (\ref{C2}), 
is easily verified to be valid.
For a finite value of the $\widetilde{y}$-coordinate and at asymptotically 
large distances $\widetilde{x}$ from the wall, we have 
\begin{equation} \label{nasymp}
n(x,y) \mathop{\sim}_{x\to\infty} 
\frac{2 c^2\ell_{\rm B}}{\pi} \frac{1}{x^4} ,
\end{equation}
which is thus $y$-independent.
This asymptotic relation is partially nonuniversal since it contains 
the surface charge parameter $c$, however it does not involve the parameter
$a$.
Since
\begin{equation}
(-e) \ell_{\rm B}^2 \int_{-\infty}^{\infty} {\rm d}\widetilde{y} 
\int_0^{\infty} {\rm d}\widetilde{x} n(\widetilde{x},\widetilde{y}) 
= \frac{(-e)}{\ell_{\rm B}} \left( \frac{a}{\sqrt{a^2-c^2}} -1 \right) 
\end{equation}
and
\begin{equation}
\qquad e \ell_{\rm B} \int_{-\infty}^{\infty} 
{\rm d}\widetilde{y} \sigma(\widetilde{y}) 
= \frac{e}{\ell_{\rm B}} \frac{a}{\sqrt{a^2-c^2}} ,
\end{equation}
one particle (per unit length in the $z$ direction) is evaporated in
the sense of the Manning-Oosawa condensation 
\cite{Oosawa68,Manning69,BuOr06,Juan}.

Other solutions are given by the choices 
\begin{equation}
Y_1(z) = z^n \quad (n=2,3,\ldots) , \qquad Y_2(z) = 1 .
\end{equation}
They possess qualitatively the same features as the $n=1$ case.
In particular, the asymptotic decay of the density profile is of the type
$n(x,y) \sim x^{-2-2n}$ for $x\to\infty$.
We do not dwell further on this family for the following reason. 
While we started the analysis in planar geometry, the very form of 
the solutions obtained (see Eq. \eqref{eq:phisol}), together with 
the intrusion of a Manning-like evaporation phenomenon, 
indicates that we are actually contemplating the potential created by 
a charged cylinder, and that cylindrical coordinates with radial variable 
$\widetilde r=\sqrt{(\widetilde x +a)^2 + \widetilde y^2}$
would simplify the formulation, for the angular dependence is here absent.
The resulting charged cylinder problem is thus isotropic (homogeneous surface 
charge), and the charge inhomogeneity obtained with Cartesian coordinates is 
thus artificial, stemming from an inappropriate choice of coordinates.
Besides, the solution \eqref{eq:phisol} actually corresponds to a
non-neutral system where, beyond the unavoidable Manning evaporation 
\cite{Oosawa68,Manning69,BuOr06,Juan}, there are too few counterions 
to neutralize the cylinder charge. 

All these solutions (including $n=1$) correspond to ``initially non-neutral'' 
cylindrical geometry configurations, since the potential decays too fast 
at infinity; we do not have a $-2\ln r$ but a $-2(1+n)\ln r$. 
What is meant here is that beyond the unavoidable Manning evaporation 
phenomenon, the solutions here correspond to a non-neutral system enclosed 
in a concentric Wigner-Seitz cylinder, the radius of which is sent to 
infinity \cite{rque200}. 

\renewcommand{\theequation}{C.\arabic{equation}}
\setcounter{equation}{0}

\section{Homogeneous surface charge density for systems with salt}
\label{app:salthomogeneous}
Introducing the dimensionless coordinate $\widetilde{x}=\kappa x$,
the one-dimensional version of the PB equation (\ref{PB2}) is written as
\begin{equation} \label{PB2uniform}
\frac{d^2\phi}{d\widetilde{x}^2} = \sinh \phi
\end{equation}
and the boundary condition (\ref{BCreduced}) at $\widetilde{x}=0$ takes 
the form 
\begin{equation} \label{BC2uniform}
-\frac{\partial\phi}{\partial \widetilde{x}}\Big\vert_{\widetilde{x}=0} = 
\frac{4\pi \ell_{\rm B} \sigma}{\kappa} \equiv a .
\end{equation} 
The potential $\phi$ is positive and its derivative $\phi'(\widetilde{x})$
negative for all $\widetilde{x}\ge 0$ (we are dealing with a positively charged 
surface). 
Multiplying the PB equation (\ref{PB2uniform}) by $\phi'(\widetilde{x})$, 
it can be simply integrated to the one \cite{Andelman06}
\begin{equation}
\phi'(\widetilde{x}) = - 2 \sinh \frac{\phi(\widetilde{x})}{2} ,
\end{equation}
which has the explicit solution
\begin{equation} \label{homosol}
\phi =  2 \ln\left( \frac{e^{\widetilde{x}}+\xi}{e^{\widetilde{x}}-\xi}
\right) , \qquad a = \frac{4\xi}{1-\xi^2} .
\end{equation}
In order to ensure the positivity of $\phi$, the parameter $\xi$ is
chosen as the positive root of the quadratic equation, 
\begin{equation} \label{xi}
\xi = \frac{-2 + \sqrt{4+a^2}}{a} .
\end{equation}
Its value is from the interval $(0,1)$, namely $\xi\to 0$ for $a\to 0$
(small $\sigma$) and $\xi\to 1$ for $a\to \infty$ (large $\sigma$).

At large distances from the wall, $\phi$ decays to zero exponentially,
\begin{equation}
\phi \mathop{\sim}_{\widetilde{x}\to\infty} 4 \xi e^{-\widetilde{x}} ,
\end{equation} 
as it should be for dense Coulomb systems.
The species densities
\begin{equation}
n_{\pm}(\widetilde{x}) = \frac{n}{2} e^{\mp \phi(\widetilde{x})} = \frac{n}{2}
\left( \frac{1 \mp \xi e^{-\widetilde{x}}}{1 \pm \xi e^{-\widetilde{x}}}
\right)^2 
\end{equation}
also decay exponentially to their bulk value $n/2$, from below for coions
and from above for counterions.
The total particle density at the wall
\begin{equation}
n(0) = n_+(0) + n_-(0) = n \cosh\phi(0) = n + 2\pi\ell_{\rm B} \sigma^2
\end{equation}
fulfills the contact theorem (\ref{contact}).

The Debye-H\"uckel (DH) approach is based on the linearization of 
the PB equation (\ref{PB2uniform}),
\begin{equation} \label{DHeq}
\frac{d^2\phi_{\rm DH}}{d\widetilde{x}^2} = \phi_{\rm DH} .
\end{equation}
The regular solution of this equation with the boundary condition 
(\ref{BC2uniform}) reads as
\begin{equation} \label{DHsol}
\phi_{\rm DH} = \frac{4\pi\ell_{\rm B}\sigma}{\kappa} 
\exp\left( -\widetilde{x} \right) .
\end{equation}
Also the original nonlinear PB equation (\ref{PB2uniform}) can be linearized
at $\widetilde{x}\to\infty$ since $\phi$ is small, and the general solution
of the linearized equation is analogous to the DH one (\ref{DHeq}), up
to a $\sigma$-dependent prefactor,
\begin{equation}
\phi \mathop{\sim}_{\widetilde{x}\to\infty} 
A(\sigma) \exp\left( -\widetilde{x} \right) .
\end{equation}
In analogy with the DH solution (\ref{DHsol}), the prefactor $A(\sigma)$ 
defines an effective (or renormalized) surface charge density 
$\sigma_{\rm eff}$ via the relation \cite{Alexander84,Diehl01,Trizac02,Samaj15} 
\begin{equation} 
A(\sigma) = \frac{4\pi\ell_{\rm B}\sigma_{\rm eff}}{\kappa} .
\end{equation}
The explicit nonlinear solution (\ref{homosol}), when expanded in 
$\exp(-\widetilde{x})$, implies
\begin{equation}
\frac{4\pi\ell_{\rm B}\sigma_{\rm eff}}{\kappa} = 4 \xi .
\end{equation}
For small $\sigma$, $\xi\sim a/4$ and $\sigma_{\rm eff}\sim \sigma$.
In the limit $\sigma\to\infty$, $\xi\to 1$ and $\sigma_{\rm eff}$ saturates
to a finite value given by
\begin{equation}
\frac{\pi\ell_{\rm B}\sigma_{\rm eff}^{\rm sat}}{\kappa} = 1 .
\end{equation}


\begin{thebibliography}{11}
\bibitem{rque10}
It was argued in Ref. \cite{Landy10} that in the weak coupling regime, 
counterion surface enhancement and pressure reduction can facilitate 
both overcharging and like-charge attraction.
Studies of surfaces with quenched charge disorder in the strong coupling 
regime \cite{Naji05,Ghodrat15,Bakhshandeh15} indicate that 
the disorder-induced attraction between the charged surfaces can become much 
stronger than their van der Waals interaction.
\bibitem{rque102}
Note that while the period of charge modulation along the plate is arbitrary 
when compared to the Debye length, the screening length perpendicular to 
the plate is smaller, or at most equal to the Debye length.
\bibitem{rque200} 
We note in passing that while it is possible to find solutions that behave 
in $-2 \xi \log \widetilde r$ at large distances with $\xi>1$, cases
with $\xi<1$ are precluded since they correspond to a situation where all 
counterions would evaporate to infinity, by the Manning mechanism.
Besides, as emphasized in the main text, the most interesting
solutions behave at large distances as  $-2 \log \widetilde r$.
\bibitem{comment2001} 
\emma{A comment is in order here, formulated 
in the planar case for simplicity. Starting from one of our solutions,
one may compute the normal electric field at an arbitrary distance on a parallel 
plane, and infer from this the corresponding surface charge density. 
The potential outside the two plates will be unaffected and coincide 
with that of our initial  
single plate problem; also unaffected will be the total pressure, which has to be the same 
on both plates, and always vanishing in our case. This is a way to extend our one-plate results 
to two-plate interactions. Such a route, however, can be seen as somewhat artificial: 
we cannot choose the charge pattern on the second plate, and the net pressure 
will always be zero.}
\end{thebibliography}
\end{document}